\documentclass[prb,showpacs,twocolumn,superscriptaddress,floatfix,longbibliography,nofootinbib]{revtex4-2}
\usepackage{graphicx,amsfonts,amssymb,amsmath,xspace,dsfont}
\usepackage[colorlinks=true,citecolor=green,linkcolor=red,urlcolor=blue]{hyperref}
\usepackage{subfigure}
\usepackage{blindtext}

\usepackage{color}
\definecolor{g}{rgb}{.1,0.4,.1} 
\definecolor{b}{rgb}{0,0.2,1}
\definecolor{rouge}{rgb}{0.82,0.,0.}
\definecolor{vert}{rgb}{0.,0.82,0.}
\definecolor{orange}{rgb}{1,0.5,0.}
\definecolor{bleu}{rgb}{0.,0.,0.82}
\definecolor{m}{rgb}{0.82,0.,0.82}
\definecolor{vert2}{rgb}{0.,0.5,0.}
\definecolor{rougeclair}{rgb}{1.0,0.7,0.7}

\newcommand{\red}{\color{rouge}}

\newcommand{\beq}{\begin{equation}}
\newcommand{\be}{\begin{equation}}
\newcommand{\beqn}{\begin{eqnarray}}
\newcommand{\eeq}{\end{equation}}
\newcommand{\ee}{\end{equation}}
\newcommand{\eeqn}{\end{eqnarray}}

\begin{document}

\title{Density of states of tight-binding models in the hyperbolic plane}

\author{R\'emy Mosseri}
\email{remy.mosseri@upmc.fr}
\affiliation{Sorbonne Universit\'e, CNRS, Laboratoire de Physique Th\'eorique de la Mati\`ere Condens\'ee, LPTMC, 75005 Paris, France}

\author{Julien Vidal}
\email{vidal@lptmc.jussieu.fr}
\affiliation{Sorbonne Universit\'e, CNRS, Laboratoire de Physique Th\'eorique de la Mati\`ere Condens\'ee, LPTMC, 75005 Paris, France}

\begin{abstract}
We study the energy spectrum of tight-binding Hamiltonians for regular hyperbolic tilings. More specifically, we compute the density of states using the continued-fraction expansion of the Green's function on finite-size systems with more than $10^9$ sites and open boundary conditions.  The coefficients of this expansion are found to quickly converge, so that the thermodynamic limit can be inferred quite accurately. This density of states is in stark contrast with the prediction stemming from  the recently proposed hyperbolic band theory. Thus, we conclude that the fraction of the energy spectrum described by the hyperbolic Bloch-like wave eigenfunctions vanishes in the thermodynamic limit. 
\end{abstract}

\maketitle

%
%
\section{Introduction } 
%
%

Since the early days of quantum mechanics, the study of electronic properties of crystalline solids has been an evergrowing field of research. In particular, the celebrated Bloch's theorem~\cite{Bloch29},  anticipated by Floquet~\cite{Floquet83} in 1883, has given rise to the band theory which is at the heart of most current electronic devices. The band theory essentially originates from the regular arrangement of atoms in solids that are classified, geometrically, by their symmetry group.  In the two-dimensional (2D) Euclidean plane (flat curvature), all periodic tessellations can be constructed from five Bravais lattices and 17 wallpaper groups. Importantly, the translation group associated with the Bravais lattice is Abelian and its 1D irreducible representations (irreps) may be seen as the cornerstone of Bloch waves.  

By contrast, in the hyperbolic plane $\mathbf{H}^2$ (constant negative curvature), there are infinitely-many regular tilings characterized by their Coxeter reflection group~\cite{Coxeter73,Magnus74}. Recently, Maciejko and Rayan proposed to use the translation Fuchsian group $\Gamma$ which is a subgroup of the Coxeter reflection group to build  the counterpart of Bloch waves in the hyperbolic plane~\cite{Maciejko21,Maciejko22} (see also Ref.~\cite{Boettcher22}). However, since $\Gamma$ is a noncommutative group, it does not admit only 1D irreps so that such an approach, dubbed  hyperbolic band theory (HBT),  also requires that higher-dimensional irreps~\cite{Maciejko22,Cheng22} be considered.

An important open question is therefore to determine the relative weight of the different irreps of $\Gamma$. In this paper, we address this issue by considering regular hyperbolic tilings for which we compute the density of states (DOS) of a tight-binding Hamiltonian. We focus on a specific set of hyperbolic tilings but our approach, based on the continued-fraction method, can equally be applied to any regular tiling. In Sec.~\ref{sec:model}, we briefly recall some basic properties of these tilings, and we introduce the model. Section~\ref{sec:cont_frac} provides a short pedagogical introduction to the continued-fraction method and explains how a rapid convergence of the coefficients allows for a precise determination of the DOS which are discussed in Sec.~\ref{sec:nde}.

By comparing the full DOS with the one coming from the Abelian hyperbolic band theory (AHBT) based on 1D irreps of $\Gamma$ (see Sec.~\ref{sec:comp}), we conclude that {\it the fraction of the full spectrum captured by the AHBT vanishes in the thermodynamic limit}. Appendix~\ref{app:size} gives informations about the shell-by-shell construction of the clusters, and Appendix~\ref{app:coeff} gives the list of coefficients used to compute the DOS.

%
%
\section{Tilings and Model}
\label{sec:model}
%
%

Two-dimensional regular tilings made of $p$-gons (polygons with $p$ sides) and $q-$fold coordinated sites are denoted by the Schl\"afli symbol $\{p,q\}$~\cite{Coxeter73}. When \mbox{$(p-2)(q-2)>4$}, these tilings can be embedded in the negatively curved hyperbolic plane $\mathbf{H}^2$. When \mbox{$(p-2)(q-2)=4$}, one recovers the usual square $\{4,4\}$, triangular $\{3,6\}$, and honeycomb $\{6,3\}$ lattices that are the only regular tilings of the flat Euclidean plane. Finally, when \mbox{$(p-2)(q-2)<4$}, one gets the five Platonic solids, namely, the tetrahedron $\{3,3\}$, the cube $\{4,3\}$,  the octahedron $\{3,4\}$, the  dodecahedron $\{5,3\}$, and the icosahedron $\{3,5\}$ which can be embedded in the positively curved sphere $\mathbf{S}^2$. The full symmetry group of a  $\{p,q\}$ tiling is the Coxeter reflection group $[p,q]$ generated by reflections in the sides of a fundamental triangular region known as the orthoscheme~\cite{Magnus74}. 

Our main goal is to determine the DOS of the standard tight-binding Hamiltonian defined on a  $\{p,q\}$ tiling: 
%
\begin{equation}
H=- t \sum_{\langle i,j\rangle} | i \rangle  \langle j |,
\label{eq:hamiltonian}
\end{equation}
where $\langle i,j\rangle$ stands for nearest-neighbor sites and where $|i\rangle$ is a state localized on site (vertex) $i$ of the tiling. In the following, we set the energy unit $t=1$ so that $H$ is simply the opposite of the adjacency matrix. We are interested in analyzing the spectrum of $H$ in the thermodynamic limit, i.e., for an infinite tiling.

\begin{figure}[t]
\centering
\includegraphics[width=0.8\columnwidth]{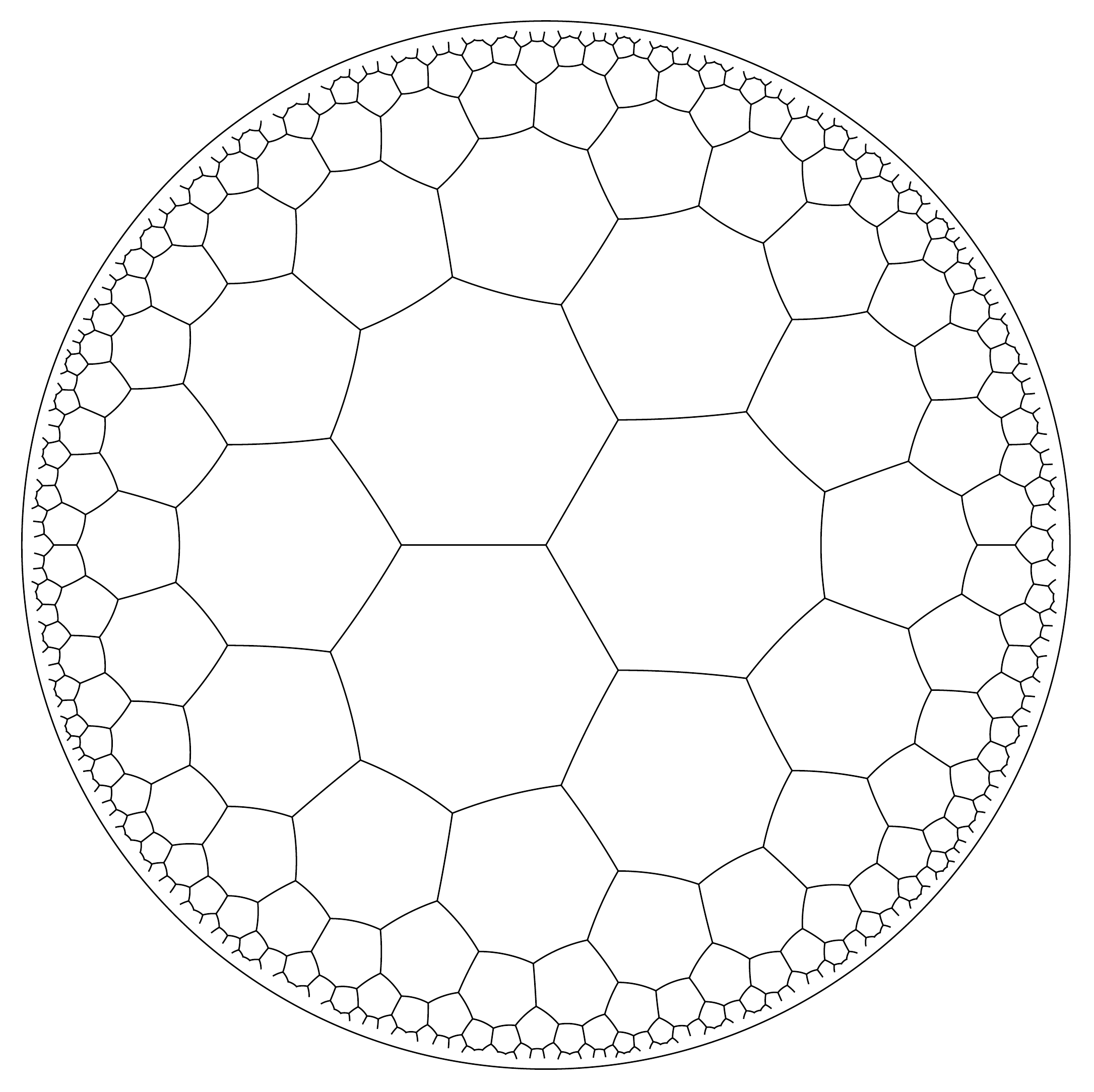}
\caption{A piece of the $\{7,3\}$ hyperbolic tiling with open boundary conditions and a radius $R=10$. It contains \mbox{472 (bulk)+ 270 (boundary) = 742 vertices}. Here, we use the standard Poincar\'e disk conformal representation of the hyperbolic plane.}
\label{fig:73OBC}
\end{figure}

A possible approach consists in performing exact diagonalizations (ED) of larger and larger clusters but, for hyperbolic tilings \mbox{[$(p-2)(q-2)>4$]}, there are several difficulties. Indeed, if one uses clusters with open boundary conditions, the ratio between the number of sites on the boundary and the number of sites in the bulk goes to a finite constant (see Fig.~\ref{fig:73OBC} for an illustration and Appendix~\ref{app:size} for a quantitative discussion) in the thermodynamic limit, whereas it vanishes in the Euclidean plane. 
%
This well-known phenomenon is due to the negative curvature of $\mathbf{H}^2$ and prevents any reliable extrapolation of the spectrum due to spurious edge states.  

To avoid boundary effects, one may alternatively consider clusters with periodic boundary conditions but another difficulty arises in this case. Indeed, the Euler-Poincar\'e characteristic $\chi$ for a compact (orientable) surface of genus $g$ reads
 %
\begin{equation}
\chi=2-2g=V-E+F,
\label{eq:EP_char}
\end{equation}
where $V$, $E$, and $F$ are the number of vertices, edges, and faces, respectively. For any hyperbolic $\{p,q\}$ tiling, one further has $pF=2E=qV$, so that one immediately gets 
%
\begin{equation}
g-1=V\, \frac{p\,q-2(p+q)}{4 p}.
\label{eq:V_vs_g}
\end{equation}
This relation shows that the genus of the surface is proportional to the number of sites, i.e., $g\propto V$.  Thus, apart from the practical difficulty in building large-genus compact systems for arbitrary $\{p,q\}$ tiling, the main problem comes from the so-called systoles defined as the shortest noncontractible loops of the periodic tiling and whose typical length scales as $\log V$~\cite{Guth10}. As a direct consequence, a finite-size cluster with $V$ vertices (sites) and periodic boundary conditions only captures the exact $n$ first moments of the spectrum of the infinite tiling with $n \propto \log V$ (see below for more details). For comparison, in the Euclidean case ($g=1$), $n \propto \sqrt{V}$. As a conclusion, although ED of periodic clusters is an efficient tool to study the tight-binding Hamiltonian for  Euclidean tilings, it is clearly doomed to failure for hyperbolic tilings due to important finite-size effects.

%
%
\section{The continued-fraction method}
\label{sec:cont_frac}
%
%

Here, we use an alternative approach to compute the DOS of the infinite-tiling spectrum. This method, known as the continued-fraction method, consists in expanding the diagonal matrix elements of the Green's function $G(E)=1/(E-H)$ as follows~\cite{Haydock72,Gaspard73,Haydock75}:  
\beq
[G(E)]_{\alpha \alpha}=
\frac{1}{E-a_1-\frac{b_1}{E-a_2-\frac{b_2}{E-a_3-\frac{b_3}{\dots}}}},
\label{eq:cfe}
\eeq
where the coefficients $(a_n,b_n)$ are rational numbers which depend on the state $|\alpha\rangle$ considered. These coefficients are directly related to those computed via the recursion method~\cite{Haydock72}.

The local density of states (LDOS) at energy $E$ associated with any state $|\alpha\rangle$ is then given by 
%
\beq
\rho_\alpha(E)=-\frac{1}{\pi} \lim_{\varepsilon \rightarrow 0^+} {\rm Im}[G(E+{\rm i} \, \varepsilon)]_{\alpha \alpha},
\label{eq:nde_cf}
\eeq
%
so that
%
\beq
\int_{-\infty}^{+\infty} \rho_\alpha(E) \, {\rm d}E=1.
\label{eq:norm}
\eeq
%
Since, for regular tilings, all sites are equivalent, the LDOS associated with a site $i$, $\rho_i (E)$,  is the same as the total DOS (up to a normalization factor). Thus, the problem amounts to computing the coefficients $(a_n,b_n)$ starting from an initial state located on a site $i$. These coefficients are directly related to the moments of the LDOS. More precisely, computing $n$ coefficients gives access to the first $2n$ moments of the LDOS, $\langle i| H^{m\leqslant 2n} |i \rangle$, and requires a cluster of radius $R=n$ (here, ``radius" means the shortest discrete graph path going from the center  to the boundary).  For instance, the cluster shown in Fig.~\ref{fig:73OBC} allows one to compute the first ten coefficients. For bipartite tilings, one has $a_{n\geqslant 1}=0$ which is reminiscent of the fact that $\langle i| H^{2m+1} |i \rangle=0$ for all $m \in \mathbb{N}$.
%
%
\begin{figure}[t]
\centering
\includegraphics[width=0.7\columnwidth]{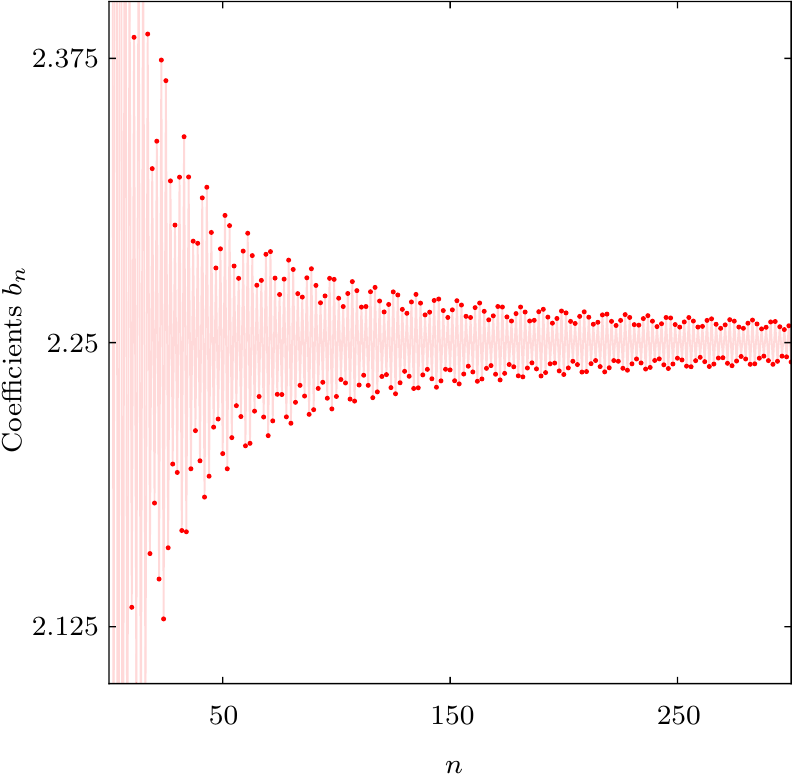}
\caption{The first 300 continued-fraction coefficients $b_n$ associated with a single site of the honeycomb lattice computed on a cluster with $V=135\,451$ sites. Because of the Van Hove singularities, the coefficients (slowly) converge with oscillations towards the asymptotic value $b_\infty=9/4$.The red line is a guide for the eyes.}
\label{fig:coeff6}
\end{figure}
%
%
The large-$n$ limit of $(a_n,b_n)$ depends on properties of the DOS. Importantly, if these coefficients converge towards unique values $(a_\infty,b_\infty)$ then the DOS is gapless. Furthermore, if the DOS contains Van Hove singularities, oscillations are expected~\cite{Hodges77}. As an example, we show in Fig.~\ref{fig:coeff6} the first 300 coefficients $b_n$ of the honeycomb tiling. The slow convergence towards the asymptotic value $b_\infty=9/4$ is due to a vanishing DOS at $E=0$, whereas oscillations originates from the two well-known Van Hove singularities at $E=\pm 1$ (see Fig.~\ref{fig:DOS_main_6_B} left panel).
By contrast, when the DOS is smooth and gapless, one expects a fast convergence of the coefficients as is the case, for instance, in the $3-$regular Bethe lattice which corresponds to the $\{\infty,3\}$ tiling~\cite{Mosseri82} and for which one gets $a_{n\geqslant 1}=0$, $b_1=3$, and $b_{n\geqslant 2}=2$ (see Fig.~\ref{fig:DOS_main_6_B}, right panel, for the DOS).

These considerations lead us to discuss the termination of the continued fraction. If the coefficients $(a_n,b_n)$ converge for sufficiently large $n$, one can replace them beyond a given $n$,  by their extrapolated asymptotic values $(a_\infty,b_\infty)$. This approximation can be interpreted as embedding the cluster under consideration into an effective medium, hence suppressing spurious edges states. Then, introducing the fraction termination 
%
%
\beq
t(E)=\frac{1}{E-a_\infty - b_\infty t(E)},
\label{eq:termination}
\eeq
%
%
i.e.,
%
%
\beq
t(E)=\frac{1}{2 b_\infty} \Big[ E-a_\infty - \sqrt{(E-a_\infty)^2-4 \, b_\infty} \Big], 
\label{eq:termination2}
\eeq
%
%
one can obtain a very good approximation of the DOS and check its convergence by increasing the value of $n$ beyond which we used the asymptotic values. 
Moreover, using Eqs.~(\ref{eq:cfe}), (\ref{eq:nde_cf}), and (\ref{eq:termination2}), one finds a nonvanishing DOS only when $E\in[E_-,E_+]$, where 
%
%
\beq
E_\pm=a_\infty \pm 2 \, \sqrt{b_\infty}.
\label{eq:boundaries}
\eeq 
%
%
For the two cases discussed above, one recovers the well-known upper and lower bounds of the honeycomb lattice~\cite{Hobson53} (\mbox{$E_\pm=\pm 3$}), as well as for the $3$-regular Bethe lattice~\cite{Thorpe81}  \mbox{($E_\pm=\pm 2 \, \sqrt{2}$)}. For these tilings, we checked explicitly that, whenever present, all singularities in the Green's function lie in the interval $[E_-,E_+]$, i.e., 
%
%
\beq
\int_{-\infty}^{+\infty} \rho_i(E) \, {\rm d}E  =\int_{E_-}^{E_+} \rho_i(E) \, {\rm d}E =1.
\label{eq:norm}
\eeq
%
%
However, let us stress that this would be different if the spectrum of $H$  would contain isolated flat bands with a finite spectral weight as, for instance, in the Kagome-like hyperbolic tilings  discussed in Refs.~\cite{Kollar19,Kollar20,Bzdusek22,Mosseri22}. In this case, extra poles would exist in the Green's function. 

%
%
\section{Density of states of $\{p,3\}$ tilings}
\label{sec:nde} 
%
%

In this paper, we focus on hyperbolic $\{p,3\}$ tilings, and we used the continued-fraction method to compute the DOS of these tilings. Because $\mathbf{H}^2$ is negatively curved, the number of sites in a cluster of typical radius $R$ grows much faster than in the Euclidean case (${\rm e}^{\#R}$ instead of  $R^2$), as shown in Appendix~\ref{app:size}. This constitutes a strong limitation in the calculations of the continued-fraction coefficients. Furthermore, the curvature increases with $p$, so that, for a given radius $R$ which determines the maximum number of  computable coefficients, the number of sites of the corresponding cluster also increases with $p$. Here, we typically used a value of $R$ which leads to clusters \mbox{with $\sim 10^9$ sites} (see Table~\ref{tab:data} for details).
%
%
\begin{table}[b]
\center
\begin{tabular}{| c | c | c | c| c| c| c|}
\hline
$p$ & $R$ & $V$ & $a_{\infty}$ & $b_{\infty}$ & $E_-$ & $E_+$ \\
\hline
7 & 42 & 1 054 313 137 & -0.1795(1) & 1.9066(6)& -2.9411 & 2.5821\\
\hline
8 & 35 & 1 049 446 747 & 0 & 2.1095(4)& -2.9048 & 2.9048\\
\hline
9& 32& 1 165 124 974 & -0.0808(2) & 1.9606(3)& -2.8812 & 2.7196\\
\hline
10 & 31 & 1 342 655 086 & 0 & 2.0528(3)& -2.8656 &  2.8656\\
\hline
11 & 30 & 1 279 395 802 & -0.0368(1) & 1.9851(2)& -2.8547 & 2.7811\\
\hline
12 & 30 & 1 675 149 250 & 0 & 2.0266(3)&-2.8471& 2.8471\\
\hline
\end{tabular}
\caption{
For the $\{p,3\}$ tilings studied in this paper, this table gives the radius $R$ of the clusters,  the corresponding number of sites $V$ (see Appendix~\ref{app:size}), the asymptotic coefficients ($a_{\infty},b_{\infty}$) extrapolated from the data given in Appendix~\ref{app:coeff}, and the corresponding boundaries of the energy range where the DOS is nonvanishing [see Eq.~(\ref{eq:boundaries})].
}
\label{tab:data}
\end{table}
%
%
Computing  $n$  continued-fraction coefficients  requires the adjacency matrix of the graph formed by the $R=n$ first shells surrounding a given site. Therefore, we applied the recursion algorithm on clusters built shell by shell.  The only limitation to compute more coefficients comes from the memory needed to store the Hamiltonian. 

%
%
\begin{figure}[t]
\includegraphics[width=0.7\columnwidth]{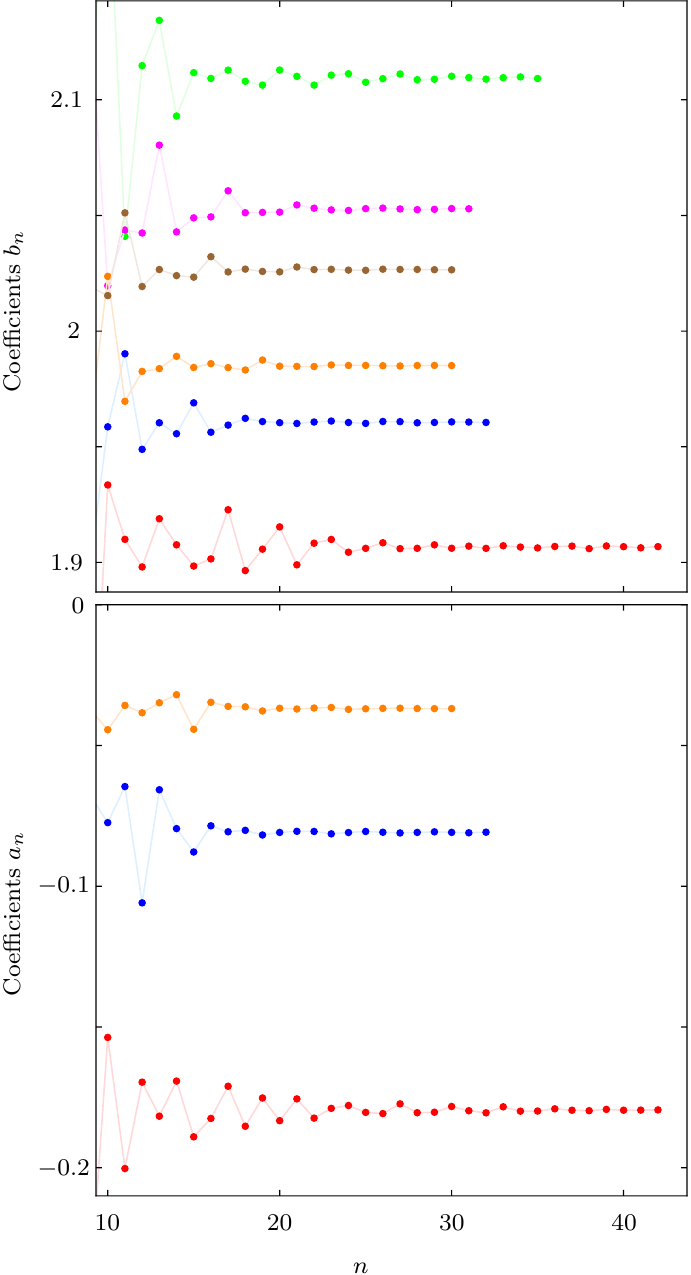}
\caption{Continued-fraction coefficients associated with a single site of hyperbolic $\{p,3\}$ tilings (see Appendix~\ref{app:coeff} for data), plotted for $n \geq 10$. Colored points: red ($p=7$), green ($p=8$), blue ($p=9$), magenta ($p=10$), orange ($p=11$), brown ($p=12$). Lines are guides for the eyes. Coefficients $a_{n \geqslant 1}=0$ for even $p$ (bipartite lattice).}
\label{fig:coeff}
\end{figure}
%
%

When considering the LDOS of $H$ for a single site, the coefficients  $(a_n,b_n)$ are rational numbers. These coefficients are given in Appendix~\ref{app:coeff} and plotted in Fig.~\ref{fig:coeff}. As can be seen, for each tiling considered, they do converge towards a unique value way faster than for the  honeycomb lattice (see Fig.~\ref{fig:coeff6} for comparison). As explained above, this indicates the absence of Van Hove singularities and of gaps in the DOS. Furthermore, this convergence allows one to extrapolate the asymptotic values $(a_{\infty},b_{\infty})$ and to compute $E_\pm$ with a better precision than with ED results~\cite{Boettcher20,Kollar20}.

%
%
\begin{figure*}[t]
\centering
\includegraphics[width=0.95\textwidth]{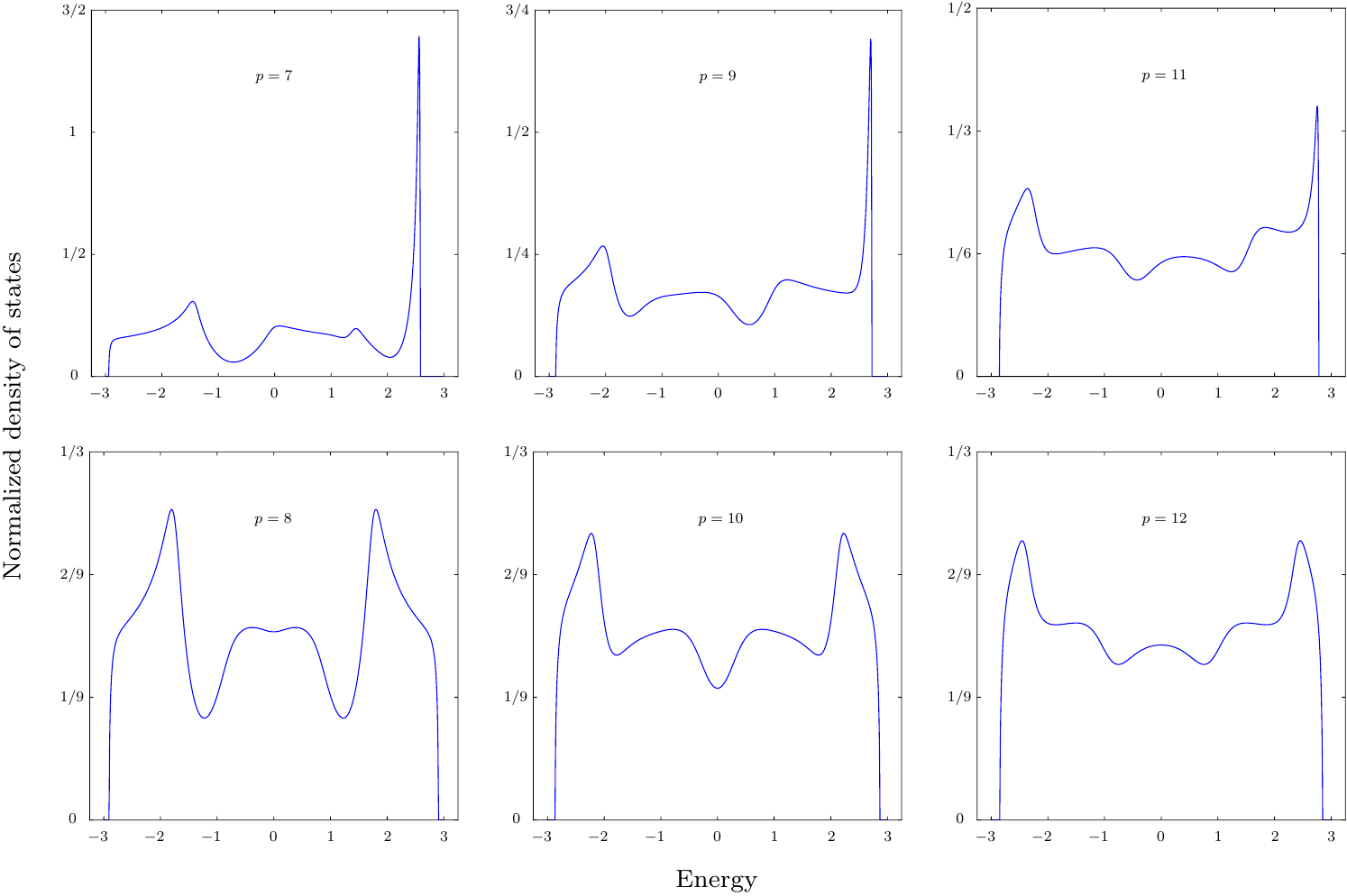}
\caption{Normalized density of states of hyperbolic $\{7\leqslant p \leqslant 12,3\}$ tilings.}
\label{fig:DOS}
\end{figure*}
%
%
%

Up to a normalization factor, the DOS in the thermodynamic limit of hyperbolic $\{p,q\}$ tilings can be defined as the quantity which has the same moments of order $m$ as the one of the Hamiltonian computed from the LDOS of a site which is the center of a cluster of radius $R\geqslant m$, for arbitrary large $m$. However, even with very large clusters, the number of exact moments (equivalently of continued fraction coefficients) remains rather small. 
Although it is hard to provide some accurate error bars, the observed fast convergence of the coefficients indicates that the large$-m$ moments are well captured by completing the continued-fraction with the asymptotic coefficients $(a_{\infty},b_{\infty})$.

Using these coefficients and the fraction termination $t(E)$, one can compute the DOS of hyperbolic $\{p,3\}$ tilings. As can be seen in Fig.~\ref{fig:DOS} for $p=7,...,12$, these DOS display several interesting features. For even (odd) $p$ , the DOS is symmetric (not symmetric) with respect to 0. This is simply due to the fact that $\{p,3\}$ tilings are (non-)bipartite for even (odd) $p$. As anticipated from the behavior of the coefficients~\cite{Hodges77}, let us stress that the peaks observed in the vicinity of $E_+$ for odd $p$ are not Van Hove singularities. We carefully checked that the DOS are finite in this energy range.

\begin{figure}[t]
\centering
\includegraphics[width=\columnwidth]{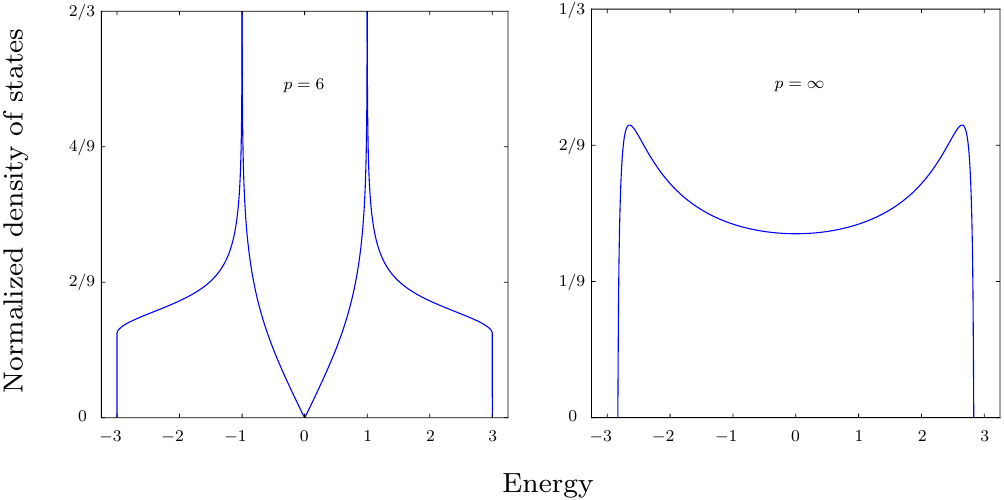}
\caption{Normalized density of states of the honeycomb  lattice  (left) and the 3-regular Bethe lattice (right).}\label{fig:DOS_main_6_B}
\end{figure}

These DOS clearly differ from the DOS of the honeycomb lattice ($p=6$)~\cite{Hobson53} 
\beq
\rho^{\{6,3\}}(E)=\frac{|E|}{\pi^2} \frac{1}{\sqrt{Z_0}}
   K\left(\sqrt{\frac{Z_1}{Z_0}} \right),
   \label{dos2.eqn}
\eeq
with
\beq
   Z_0=\left\{
   \begin{array}{ll}
      \left(1+|E|\right)^2-\frac{\left(E^2-1\right)^2}{4}, & {\rm for} \:\: |E|<1\\
      4|E|,					& {\rm for}  \:\:1\le |E|\le 3
   \end{array}
   \right.,
\eeq
and 
\beq
   Z_1=\left\{
   \begin{array}{ll}
       \left(1+|E|\right)^2-\frac{\left(E^2-1\right)^2}{4}, & {\rm for}  \:\:1\le |E|\le 3\\
            4|E|,	& {\rm for}  \:\: |E|<1\\
      \end{array}
         \right.,
\eeq
where $K$ is the complete elliptic integral of the first kind. However, when $p$ increases, these DOS converge towards the DOS of the 3-regular Bethe lattice ($p=\infty$), which reads~\cite{Thorpe81}
%
%
\beq
\rho^{\{\infty,3\}}(E)=\frac{3}{2 \pi} \frac{\sqrt{8-E^2}}{9-E^2}.
\eeq
%
%
These two (well-known) limiting cases are reproduced in Fig.~\ref{fig:DOS_main_6_B}.
%
%
%
%
The DOS displayed in Fig.~\ref{fig:DOS} must be considered as a very good approximation of the exact DOS of the infinite $\{p,3\}$ tiling, in the sense that it has the same $2R$ moments. Although it is difficult to give some error bars within the continued-fraction framework, the main source of errors comes from substituting the coefficients $(a_n,b_n)$ by their extrapolated asymptotic values $(a_{\infty},b_{\infty})$, for $n>R$. As can be checked in the data given in Appendix~\ref{app:coeff}, the relative error is $\sim10^{-4}$ (see Table~\ref{tab:data}) so that we obtain a very good approximation of the exact DOS.

The DOS of several $\{p,3\}$ tilings have recently been computed by ED of clusters with open and periodic boundary conditions. In Ref.~\cite{Kollar20}, Koll\'ar {\it et al.}  focused on $p=7, 8$, and used an arbitrary bin width to compute the DOS [see Figs.~14(a)-14(d) of Ref.\cite{Kollar20}]. In Ref.~\cite{Urwyler22}, Urwyler {\it et al.} performed a similar study for $p=8$ but used an additional filtering procedure to get rid of boundary effects together with an arbitrary Gaussian smearing function [see Fig.~1(b) of Ref.~\cite{Urwyler22}].  Some results for  $p=8,10,12$ can also be found in Ref.~\cite{Gluscevich23}, where a classification is proposed. A comparison with our results shows that ED of hyperbolic finite-size clusters with a few thousand sites can hardly reproduce the main pattern of the asymptotic DOS shown in Fig.~\ref{fig:DOS} and this comparison sheds light on the importance of boundaries for hyperbolic tilings.


%
%
\section{Comparison with hyperbolic band theory} 
\label{sec:comp} 
%
%
In the previous section, we computed the full DOS of some $\{p,3\}$ tilings. As explained above, these DOS share, by construction, the same first $2n$ moments as those of the corresponding infinite tiling, where $n$ is the number of continued-fraction coefficients computed. However, at this stage, it is important to specify what is meant by``infinite tiling". As for Euclidean tilings, the``infinite" limit of hyperbolic tilings can be obtained with either open or periodic boundary conditions by increasing the linear system size. However, in the hyperbolic case, several compactifications can be considered giving rise to completely different DOS. 
Hence, to compare our results with the predictions stemming from the AHBT, we shall first discuss the case of the infinite $\{p,q\}$ tiling which is a tessellation of the infinite hyperbolic plane $\mathbf{H}^2$ and, in a second step, the compact case.

\subsection{The infinite $\{p,q\}$ tiling}
As mentioned in Sec.~\ref{sec:model}, the symmetry group of the infinite $\{p,q\}$ tiling is the Coxeter reflection group $[p,q]$. This group contains a torsion-free Fuchsian subgroup $\Gamma$, which  describes the noncommutative translations of $\mathbf{H}^2$. Although non-Abelian, $\Gamma$ has 1D irreps that allow one to compute some eigenvalues associated with Bloch-like eigenstates~\cite{Maciejko21,Boettcher22,Maciejko22}. The AHBT aims at describing the band structure associated with these irreps. 

In the Euclidean plane, the translation group is Abelian and, hence, all irreps are 1D. Thus, the whole spectrum of $H$ can be described by the standard Bloch band theory. By contrast, in the hyperbolic plane, the weight $\omega_1$ of 1D irreps at the heart of the AHBT has been the topic of recent studies~\cite{Maciejko21,Maciejko22,Cheng22,Bzdusek22} and, to our knowledge, is still unknown. As we shall now argue, this weight is actually vanishing in the infinite $\{p,q\}$ hyperbolic tiling.
Although the irreps decomposition of an infinite discrete Fuchsian group is a complicated subject, the full DOS can always be formally decomposed as:
%
%
\beq  
\rho^{\rm full}(E)=\sum_d \omega_{d}  \, \rho^{(d)}(E),
\label{eq:irreps}
\eeq 
%
%
where $\rho^{(d)}$ is the normalized DOS obtained from all $d$-dimensional irreps of $\Gamma$ and where $\omega_{d}$ is the weight of all these representations in the decomposition of $\Gamma$ into irreps. Our goal is to evaluate $\omega_{1}$ in the thermodynamic limit.

\begin{figure}[t]
\includegraphics[width=0.7\columnwidth]{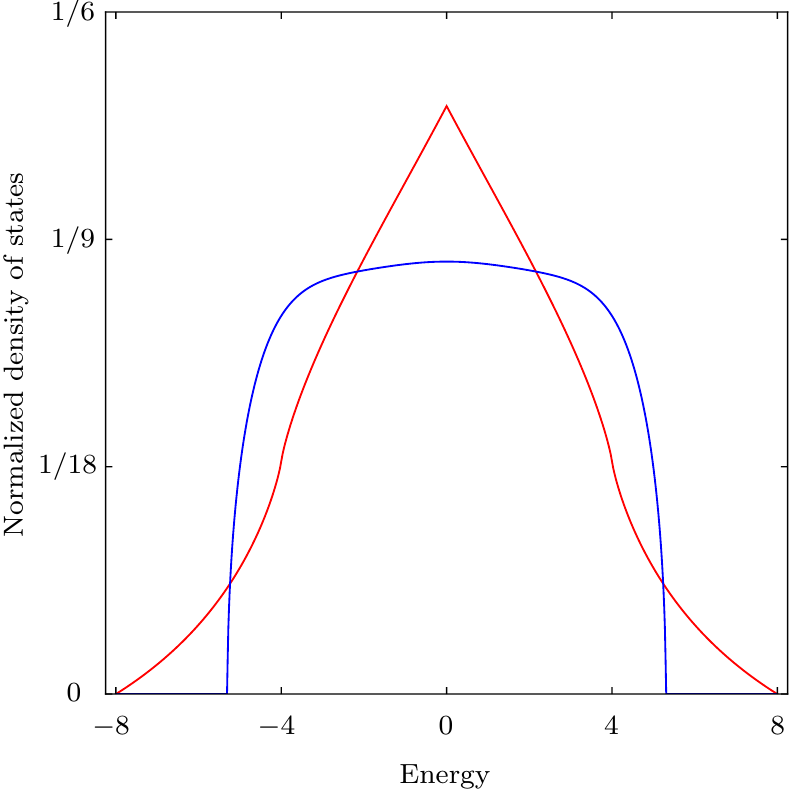}
\caption{Comparison between  the exact AHBT DOS (red) and the full DOS of the hyperbolic $\{8,8\}$ tiling computed with the continued-fraction method (blue).}
\label{fig:HBTspectra}
\end{figure}

To do so, let us focus on the hyperbolic $\{8,8\}$ tiling for which the AHBT has been developed in Ref.~\cite{Maciejko22}, but the same line of reasoning is straightforwardly adaptable to any $\{p,q\}$ tiling.  The AHBT theory for the $\{8,8\}$ tiling states that the spectrum originating from the 1D irreps of $\Gamma$ is given by
%
%
\beq
E({\bf k})= -2 \sum_{j=1,4}\cos k_j,
\label{eq:4dcubic}
\eeq
%
%
where ${\bf k}=(k_1,k_2,k_3,k_4)$ is a 4D vector whose components $k_j$ are associated with the four generators $\gamma_j$ of $\Gamma$~\cite{Maciejko22}. 
This dispersion relation is actually the same as the one of the 4D hypercubic lattice. Here, following Ref.~\cite{Maciejko22}, we consider the thermodynamic limit and assume that these momenta can take any value in the 4D first Brillouin zone, i.e.,$-\pi \leqslant k_j < \pi$.  Thus, the corresponding DOS is given by: 
%
%
\beq  
\rho^{(1)}(E)= \frac{1}{\pi} \int_0^\infty J_0( 2 u)^4 \cos(u E)\, {\rm d}u, 
\label{eq:rho1}
\eeq 
%
%
where $J_0$ is the Bessel function of the first kind. This DOS is  plotted in Fig.~\ref{fig:HBTspectra} (red line) and is nonvanishing for $E \in [-8,+8]$.

To compute the full DOS $\rho^{\rm full}(E)$ of the hyperbolic $\{8,8\}$ tiling, we use the continued-fraction method described in Sec.~\ref{sec:cont_frac}. For this tiling, the radius of largest cluster considered here is $R=10$, but, as can be inferred from Appendix~\ref{app:coeff}, we observe (again) a quick convergence of the coefficients $b_n$ that allows one to extrapolate the asymptotic value $b_\infty=7.02912(1)$. Using this value for the fraction termination, we can compute the DOS of the hyperbolic $\{8,8\}$ tiling which is nonvanishing for $E\in [E_-,E_+]$ with $E_+=-E_-=2\sqrt{b_\infty} \simeq 5.3025$. Note that our estimate of $E_+$ lies within the sharp interval $8 \times [0.662772,0.662816]$~\cite{Nagnibeda99,Gouezel15}. Furthermore, with the ten coefficients given in Appendix~\ref{app:coeff}, one can straightforwardly computes the first 20 moments of the DOS.  We  checked that these moments match with the ones given in Ref.~\cite{Lux22}, where the first 8 moments have been computed on {\it ad hoc} clusters with periodic boundary conditions (see Appendix ~\ref{app:coeff}).

%
%
%
%

As can be seen in Fig.~\ref{fig:HBTspectra} where we plotted $\rho^{(1)}$ and $\rho^{\rm full}$, there is an extended energy region where $\rho^{(1)}$ is finite and where $\rho^{\rm full}$ is vanishing,$[-8,E_-]$ (and its symmetric counterpart, $[E_+,8]$). Using Eq.~(\ref{eq:rho1}), one can compute the integrated DOS in this region
%
%
\beqn  
\int_{-8}^{E^-} \rho^{{\rm full}}(E)\,  {\rm d} E &=&0,\\
\int_{-8}^{E^-} \rho^{(1)}(E)\,  {\rm d} E &\simeq& 0.030186,
\label{eq:rho1}
\eeqn 
%
%
which, according to Eq.~(\ref{eq:irreps}), straightforwardly implies \mbox{$\omega_1=0$}. In other words, {\em the spectral weight captured by the AHBT  is vanishing in the thermodynamic limit}.

\begin{figure}[t]
\includegraphics[width=0.7\columnwidth]{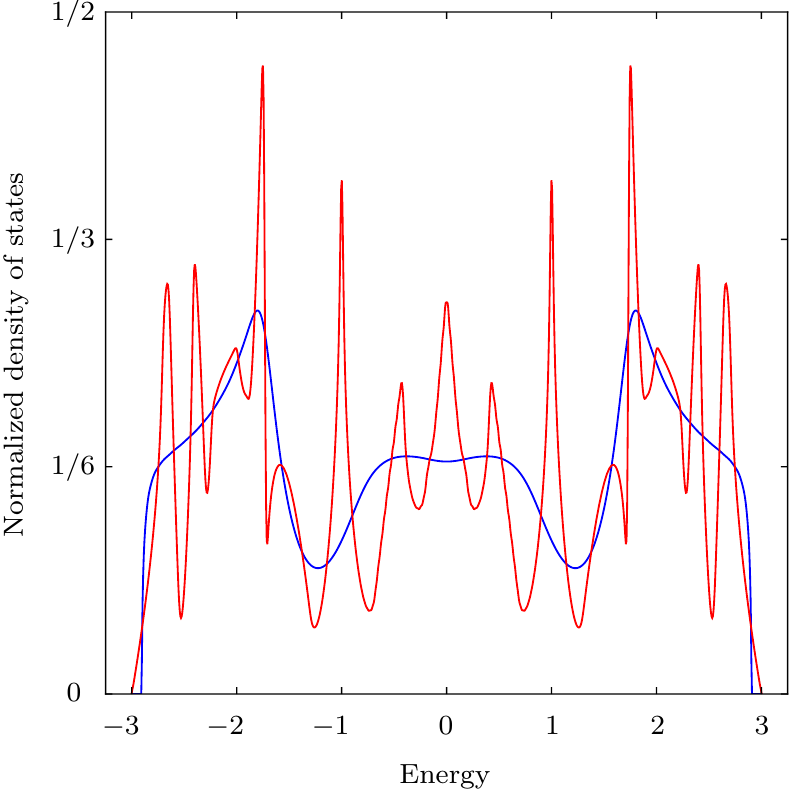}
\caption{Comparison between the numerical AHBT DOS (red) computed  with $64^4$ points in the 4D Brillouin zone (bin width=$10^{-3}$) and the full DOS of the hyperbolic $\{8,3\}$ tiling computed with the continued-fraction method (blue) also shown in Fig.~\ref{fig:DOS}. Apart from the vicinity of the sharpest peaks, we checked the convergence of the AHBT DOS by varying the number of points in the Brillouin zone and the bin width (see also Refs.~\cite{Urwyler22,Urwyler_thesis} for similar results).}
\label{fig:HBT_8_3}
\end{figure}

For a regular $\{p,q\}$ tiling, the normalized DOS is vanishing for $E \leqslant E_{-}$, where, for hyperbolic $\{p,q\}$ tiling, one has~\cite{Kollar20}
%
%
\be
-q < -q \, \sqrt{1-\alpha^2} \leqslant E_{-},
\label{eq:Cheeger}
\ee
%
%
where 
%
%
\be
\alpha=\frac{q-2}{q} \sqrt{1-\frac{4}{(p-2)(q-2)}},
\ee
%
%
is an isoperimetric constant given in Ref.~\cite{Higuchi03}, analogous to Cheeger's constant~\cite{Cheeger70}. Thus, we can conclude that, for all hyperbolic $\{p,q\}$ tiling, one has: 
%
%
\be
\int_{-q}^{-q \, \sqrt{1-\alpha^2}} \rho^{{\rm full}}(E)\,  {\rm d} E =0.
\ee
%
%
By contrast, the AHBT  leads to a nonvanishing DOS in the vicinity $E_{0}=-q$, which is always reached for ${\bf k}=0$. Indeed, for a $d$-dimensional Brillouin zone, the DOS is expected to behave as $\rho(E) \sim E^\frac{d-2}{2}$ near the band edges (see, e.g., Figs.~\ref{fig:HBTspectra} and~\ref{fig:HBT_8_3} where $d=4$), so that the integrated DOS in any finite region near $E_0$ is nonvanishing.  Hence, we conclude that $\omega_1=0$. 

As a final example, we computed the AHBT DOS for the $\{8,3\}$ tiling (see also Refs.~\cite{Urwyler22,Urwyler_thesis}) by exactly diagonalizing the ($16 \times 16$) matrix given in Ref.~\cite{Chen23}  using a discretization of the 4D Brillouin zone. As can be seen in Fig.~\ref{fig:HBT_8_3} (red), the AHBT DOS displays several well-defined peaks as well as a nonvanishing weight in the range $[-3,E_-]$ (about $0.2\%$ of the states). This again illustrates that the DOS stemming from the AHBT does not share any features with the full DOS, which is in agreement with $\omega_1=0$ but in stark contradiction with the conclusions of Ref.~\cite{Chen23}.

\subsection{The compact case}
\label{sec:PBC} 

Let us now analyze the case of  compact hyperbolic $\{p,q\}$ tilings which is discussed in details in Ref.~\cite{Maciejko22}. As discussed above, in the hyperbolic plane $\mathbf{H}^2$, these tilings are invariant under the Fuchsian group $\Gamma$. With periodic boundary conditions, the situation is different since the tilings are only invariant under the residual quotient group $G=\Gamma/\Gamma_{\rm PBC}$, where $\Gamma_{\rm PBC}$ is a finite-index normal subgroup of $\Gamma$~\cite{Maciejko22}. In this case, Eq.~(\ref{eq:irreps}) involves the irreps of $G$ and two cases must be distinguished. 

When $G$ is Abelian, the corresponding clusters, dubbed Abelian clusters in Ref.~\cite{Maciejko22}, can be fully described by the AHBT. For $p=q=8$, the full spectrum of these Abelian clusters is given by Eq.~(\ref{eq:4dcubic}) with an appropriate discretization of the 4D Brillouin zone. For these clusters, one thus has  $\omega_1=1$.  However, these Abelian clusters are locally very different from the hyperbolic $\{8,8\}$ tiling defined in $\mathbf{H}^2$ and corresponds to a  compactified version of a 4D hypercubic lattice. This is clearly seen by considering the  moments $\langle H^{m \geqslant 4} \rangle$. Indeed, for any site $i$ of the infinite $\{8,8\}$ tiling, one has $\langle i |H^{4} | i \rangle=120$, whereas, for the 4D hypercubic lattice, one gets $\langle i |H^{4} | i \rangle=168$, the difference being due to a large number of squares (4-gons) in the latter, which do not exist  in the former. We conclude that,  although the AHBT gives the full spectrum for these Abelian clusters, it does not describe the hyperbolic tilings' DOS in the thermodynamic limit (see Fig.~\ref{fig:HBTspectra}).

The second case concerns non-Abelian clusters that are associated with a non-Abelian quotient group $G$. For sufficiently large clusters, it is possible to obtain the exact moments up to a given order, but the bottleneck is then the length $l$ of the systole. However, non-Abelian $G$ has some 1D irreps. As explained after  Eq.~(\ref{eq:4dcubic}), for the compactified $\{8,8\}$ tiling, these irreps are labeled by four discrete sets of independent $k_j$ in the 4D Brillouin zone.  For each direction, the maximum number of allowed $k_j$ values is typically of order $l$, leading, at most, to $l^4$ eigenvalues (actually there are more constraints due to the high genus of the surface, which increases with the system size). As explained in Sec.~\ref{sec:model}, $l$ grows typically as  $\log V$. Since the Hilbert space dimension equals $V$, we conclude that $\omega_1$ decreases with $V$ and vanishes as $V\rightarrow\infty$. Notice that having $l$ as an upper bound in each direction is related to our consideration of clusters having increasingly correct $l$ first moments.

To conclude this section, let us stress that $G$ also has \mbox{$(d> 1)$-dimensional} irreps  labeled by a finite-dimensional discrete sets of parameters  [$(2d^2+2)$ for the $\{8,8\}$ tiling]\cite{Maciejko22}. Determining the contribution of these irreps in the full DOS, i.e., $\omega_{d>1}$,  requires a better knowledge of the corresponding non-Abelian Brillouin zone discretization as well as the constraints imposed by the systole.

%
%
%
%

%
%
\section{Conclusion}
\label{sec:ccl}
%
%
Using the continued-fraction method on large system sizes ($\sim 10^9$ sites), we computed the DOS of regular hyperbolic $\{p,3\}$ tilings for $p=7,..., 12$, which is very close to the infinite-tiling DOS (see discussion about the termination fraction in Sec.~\ref{sec:cont_frac}). These DOS are found to be smooth (no Van Hove singularities) and gapless. Importantly, we found that these DOS vanish in the  energy range $[-3,E_-]$, where $E_->-3$ satisfies Eq.~(\ref{eq:Cheeger}). This indicates that the fraction of the spectrum described by the AHBT theory for which the DOS is nonzero in the same energy range, vanishes in the thermodynamic limit. This raises important questions about the weight of higher-dimensional representations of the translation Fuchsian group $\Gamma$. In a recent work, Cheng {\it et al.}~\cite{Cheng22} considered 2D irreps of $\Gamma$ for the $\{8,8\}$ tiling. They show some cut of the corresponding 10D band structure, which extends up to the Perron-Frobenius bound $-8$. Hence, as for 1D irreps, this indicates that the weight of these 2D irreps is also very likely vanishing.  To go beyond, one definitely needs a better knowledge of the irrep decomposition of $\Gamma$, and of the associated higher-dimensional Brillouin zone geometries.
Finally, let us mention that our method can equally be applied to other kinds of Hamiltonians including complex or longer-range hoppings, multiple orbitals, etc. It can also describe gapped DOS in which case the coefficients split into subsets that converge towards different values. We hope that such a promising route could also be probed in experiments using circuit quantum electrodynamics~\cite{Kollar19}.

\acknowledgements
%
%
We thank J.-N. Fuchs, J.-P. Gaspard, S. Gou\"ezel,  and R. Vogeler for fruitful discussions. 


\begin{thebibliography}{30}%
\makeatletter
\providecommand \@ifxundefined [1]{%
 \@ifx{#1\undefined}
}%
\providecommand \@ifnum [1]{%
 \ifnum #1\expandafter \@firstoftwo
 \else \expandafter \@secondoftwo
 \fi
}%
\providecommand \@ifx [1]{%
 \ifx #1\expandafter \@firstoftwo
 \else \expandafter \@secondoftwo
 \fi
}%
\providecommand \natexlab [1]{#1}%
\providecommand \enquote  [1]{``#1''}%
\providecommand \bibnamefont  [1]{#1}%
\providecommand \bibfnamefont [1]{#1}%
\providecommand \citenamefont [1]{#1}%
\providecommand \href@noop [0]{\@secondoftwo}%
\providecommand \href [0]{\begingroup \@sanitize@url \@href}%
\providecommand \@href[1]{\@@startlink{#1}\@@href}%
\providecommand \@@href[1]{\endgroup#1\@@endlink}%
\providecommand \@sanitize@url [0]{\catcode `\\12\catcode `\$12\catcode
  `\&12\catcode `\#12\catcode `\^12\catcode `\_12\catcode `\%12\relax}%
\providecommand \@@startlink[1]{}%
\providecommand \@@endlink[0]{}%
\providecommand \url  [0]{\begingroup\@sanitize@url \@url }%
\providecommand \@url [1]{\endgroup\@href {#1}{\urlprefix }}%
\providecommand \urlprefix  [0]{URL }%
\providecommand \Eprint [0]{\href }%
\providecommand \doibase [0]{https://doi.org/}%
\providecommand \selectlanguage [0]{\@gobble}%
\providecommand \bibinfo  [0]{\@secondoftwo}%
\providecommand \bibfield  [0]{\@secondoftwo}%
\providecommand \translation [1]{[#1]}%
\providecommand \BibitemOpen [0]{}%
\providecommand \bibitemStop [0]{}%
\providecommand \bibitemNoStop [0]{.\EOS\space}%
\providecommand \EOS [0]{\spacefactor3000\relax}%
\providecommand \BibitemShut  [1]{\csname bibitem#1\endcsname}%
\let\auto@bib@innerbib\@empty
\bibitem [{\citenamefont {Bloch}(1929)}]{Bloch29}%
  \BibitemOpen
  \bibfield  {author} {\bibinfo {author} {\bibfnamefont {F.}~\bibnamefont
  {Bloch}},\ }\bibfield  {title} {\bibinfo {title} {{\"Uber die Quantenmechanik
  der Elektronen in Kristallgittern}},\ }\href
  {https://doi.org/10.1007/BF01339455} {\bibfield  {journal} {\bibinfo
  {journal} {Z. Phys.}\ }\textbf {\bibinfo {volume} {52}},\ \bibinfo {pages}
  {555} (\bibinfo {year} {1929})}\BibitemShut {NoStop}%
\bibitem [{\citenamefont {Floquet}(1883)}]{Floquet83}%
  \BibitemOpen
  \bibfield  {author} {\bibinfo {author} {\bibfnamefont {G.}~\bibnamefont
  {Floquet}},\ }\bibfield  {title} {\bibinfo {title} {{Sur les \'equations
  diff\'erentielles lin\'eaires \`a coefficients p\'eriodiques}},\ }\href
  {https://doi.org/10.1088/0022-3719/8/16/011} {\bibfield  {journal} {\bibinfo
  {journal} {Ann. Sci. de l'ENS}\ }\textbf {\bibinfo {volume} {12}},\ \bibinfo
  {pages} {47} (\bibinfo {year} {1883})}\BibitemShut {NoStop}%
\bibitem [{\citenamefont {Coxeter}(1973)}]{Coxeter73}%
  \BibitemOpen
  \bibfield  {author} {\bibinfo {author} {\bibfnamefont {H.~S.~M.}\
  \bibnamefont {Coxeter}},\ }\href@noop {} {\emph {\bibinfo {title} {{Regular
  Polytopes}}}}\ (\bibinfo  {publisher} {Dover},\ \bibinfo {address} {New
  York},\ \bibinfo {year} {1973})\BibitemShut {NoStop}%
\bibitem [{\citenamefont {Magnus}(1974)}]{Magnus74}%
  \BibitemOpen
  \bibfield  {author} {\bibinfo {author} {\bibfnamefont {W.}~\bibnamefont
  {Magnus}},\ }\href@noop {} {\emph {\bibinfo {title} {{Noneuclidean
  Tesselations and Their Groups}}}}\ (\bibinfo  {publisher} {Academic Press},\
  \bibinfo {address} {New York},\ \bibinfo {year} {1974})\BibitemShut {NoStop}%
\bibitem [{\citenamefont {Maciejko}\ and\ \citenamefont
  {Rayan}(2021)}]{Maciejko21}%
  \BibitemOpen
  \bibfield  {author} {\bibinfo {author} {\bibfnamefont {J.}~\bibnamefont
  {Maciejko}}\ and\ \bibinfo {author} {\bibfnamefont {S.}~\bibnamefont
  {Rayan}},\ }\bibfield  {title} {\bibinfo {title} {{Hyperbolic band theory}},\
  }\href {https://doi.org/10.1126/sciadv.abe9170} {\bibfield  {journal}
  {\bibinfo  {journal} {Sci. Adv.}\ }\textbf {\bibinfo {volume} {7}},\ \bibinfo
  {pages} {eabe9170} (\bibinfo {year} {2021})}\BibitemShut {NoStop}%
\bibitem [{\citenamefont {Maciejko}\ and\ \citenamefont
  {Rayan}(2022)}]{Maciejko22}%
  \BibitemOpen
  \bibfield  {author} {\bibinfo {author} {\bibfnamefont {J.}~\bibnamefont
  {Maciejko}}\ and\ \bibinfo {author} {\bibfnamefont {S.}~\bibnamefont
  {Rayan}},\ }\bibfield  {title} {\bibinfo {title} {{Automorphic Bloch theorems
  for hyperbolic lattices}},\ }\href {https://doi.org/10.1073/pnas.2116869119}
  {\bibfield  {journal} {\bibinfo  {journal} {Proc. Natl. Acad. Sci. USA}\
  }\textbf {\bibinfo {volume} {119}},\ \bibinfo {pages} {e2116869119} (\bibinfo
  {year} {2022})}\BibitemShut {NoStop}%
\bibitem [{\citenamefont {Boettcher}\ \emph {et~al.}(2022)\citenamefont
  {Boettcher}, \citenamefont {Gorshkov}, \citenamefont {Koll\'ar},
  \citenamefont {Maciejko}, \citenamefont {Rayan},\ and\ \citenamefont
  {Thomale}}]{Boettcher22}%
  \BibitemOpen
  \bibfield  {author} {\bibinfo {author} {\bibfnamefont {I.}~\bibnamefont
  {Boettcher}}, \bibinfo {author} {\bibfnamefont {A.~V.}\ \bibnamefont
  {Gorshkov}}, \bibinfo {author} {\bibfnamefont {A.~J.}\ \bibnamefont
  {Koll\'ar}}, \bibinfo {author} {\bibfnamefont {J.}~\bibnamefont {Maciejko}},
  \bibinfo {author} {\bibfnamefont {S.}~\bibnamefont {Rayan}},\ and\ \bibinfo
  {author} {\bibfnamefont {R.}~\bibnamefont {Thomale}},\ }\bibfield  {title}
  {\bibinfo {title} {{Crystallography of hyperbolic lattices}},\ }\href
  {https://doi.org/10.1103/PhysRevB.105.125118} {\bibfield  {journal} {\bibinfo
   {journal} {Phys. Rev. B}\ }\textbf {\bibinfo {volume} {105}},\ \bibinfo
  {pages} {125118} (\bibinfo {year} {2022})}\BibitemShut {NoStop}%
\bibitem [{\citenamefont {Cheng}\ \emph {et~al.}(2022)\citenamefont {Cheng},
  \citenamefont {Serafin}, \citenamefont {McInerney}, \citenamefont {Rocklin},
  \citenamefont {Sun},\ and\ \citenamefont {Mao}}]{Cheng22}%
  \BibitemOpen
  \bibfield  {author} {\bibinfo {author} {\bibfnamefont {N.}~\bibnamefont
  {Cheng}}, \bibinfo {author} {\bibfnamefont {F.}~\bibnamefont {Serafin}},
  \bibinfo {author} {\bibfnamefont {J.}~\bibnamefont {McInerney}}, \bibinfo
  {author} {\bibfnamefont {Z.}~\bibnamefont {Rocklin}}, \bibinfo {author}
  {\bibfnamefont {K.}~\bibnamefont {Sun}},\ and\ \bibinfo {author}
  {\bibfnamefont {X.}~\bibnamefont {Mao}},\ }\bibfield  {title} {\bibinfo
  {title} {{Band Theory and Boundary Modes of High-Dimensional Representations
  of Infinite Hyperbolic Lattices}},\ }\href
  {https://doi.org/10.1103/PhysRevLett.129.088002} {\bibfield  {journal}
  {\bibinfo  {journal} {Phys. Rev. Lett.}\ }\textbf {\bibinfo {volume} {129}},\
  \bibinfo {pages} {088002} (\bibinfo {year} {2022})}\BibitemShut {NoStop}%
\bibitem [{\citenamefont {Guth}()}]{Guth10}%
  \BibitemOpen
  \bibfield  {author} {\bibinfo {author} {\bibfnamefont {L.}~\bibnamefont
  {Guth}},\ }\href@noop {} {\bibinfo {title} {{Metaphors in systolic
  geometry}}},\ \bibinfo {note}
  {\href{https://doi.org/10.48550/arXiv.1003.4247}{arXiv:1003.4247}}\BibitemShut
  {NoStop}%
\bibitem [{\citenamefont {Haydock}\ \emph {et~al.}(1972)\citenamefont
  {Haydock}, \citenamefont {Heine},\ and\ \citenamefont {Kelly}}]{Haydock72}%
  \BibitemOpen
  \bibfield  {author} {\bibinfo {author} {\bibfnamefont {R.}~\bibnamefont
  {Haydock}}, \bibinfo {author} {\bibfnamefont {V.}~\bibnamefont {Heine}},\
  and\ \bibinfo {author} {\bibfnamefont {M.~J.}\ \bibnamefont {Kelly}},\
  }\bibfield  {title} {\bibinfo {title} {{Electronic structure based on the
  local atomic environment for tight-binding bands}},\ }\href
  {https://doi.org/10.1088/0022-3719/5/20/004} {\bibfield  {journal} {\bibinfo
  {journal} {J. Phys. C}\ }\textbf {\bibinfo {volume} {5}},\ \bibinfo {pages}
  {2845} (\bibinfo {year} {1972})}\BibitemShut {NoStop}%
\bibitem [{\citenamefont {Gaspard}\ and\ \citenamefont
  {Cyrot-Lackmann}(1973)}]{Gaspard73}%
  \BibitemOpen
  \bibfield  {author} {\bibinfo {author} {\bibfnamefont {J.~P.}\ \bibnamefont
  {Gaspard}}\ and\ \bibinfo {author} {\bibfnamefont {F.}~\bibnamefont
  {Cyrot-Lackmann}},\ }\bibfield  {title} {\bibinfo {title} {{Density of states
  from moments. Application to the impurity band}},\ }\href
  {https://doi.org/10.1088/0022-3719/6/21/012} {\bibfield  {journal} {\bibinfo
  {journal} {J. Phys. C}\ }\textbf {\bibinfo {volume} {6}},\ \bibinfo {pages}
  {3077} (\bibinfo {year} {1973})}\BibitemShut {NoStop}%
\bibitem [{\citenamefont {Haydock}\ \emph {et~al.}(1975)\citenamefont
  {Haydock}, \citenamefont {Heine},\ and\ \citenamefont {Kelly}}]{Haydock75}%
  \BibitemOpen
  \bibfield  {author} {\bibinfo {author} {\bibfnamefont {R.}~\bibnamefont
  {Haydock}}, \bibinfo {author} {\bibfnamefont {V.}~\bibnamefont {Heine}},\
  and\ \bibinfo {author} {\bibfnamefont {M.~J.}\ \bibnamefont {Kelly}},\
  }\bibfield  {title} {\bibinfo {title} {{Electronic structure based on the
  local atomic environment for tight-binding bands. II}},\ }\href
  {https://doi.org/10.1088/0022-3719/8/16/011} {\bibfield  {journal} {\bibinfo
  {journal} {J. Phys. C}\ }\textbf {\bibinfo {volume} {8}},\ \bibinfo {pages}
  {2591} (\bibinfo {year} {1975})}\BibitemShut {NoStop}%
\bibitem [{\citenamefont {Hodges}(1977)}]{Hodges77}%
  \BibitemOpen
  \bibfield  {author} {\bibinfo {author} {\bibfnamefont {C.~H.}\ \bibnamefont
  {Hodges}},\ }\bibfield  {title} {\bibinfo {title} {{Van Hove singularities
  and continued fraction coefficients}},\ }\href
  {https://doi.org/10.1051/jphyslet:01977003809018700} {\bibfield  {journal}
  {\bibinfo  {journal} {J. Phys. Lett.}\ }\textbf {\bibinfo {volume} {38}},\
  \bibinfo {pages} {187} (\bibinfo {year} {1977})}\BibitemShut {NoStop}%
\bibitem [{\citenamefont {Mosseri}\ and\ \citenamefont
  {Sadoc}(1982)}]{Mosseri82}%
  \BibitemOpen
  \bibfield  {author} {\bibinfo {author} {\bibfnamefont {R.}~\bibnamefont
  {Mosseri}}\ and\ \bibinfo {author} {\bibfnamefont {J.~F.}\ \bibnamefont
  {Sadoc}},\ }\bibfield  {title} {\bibinfo {title} {{The Bethe Lattice : A
  Regular Tiling of the Hyperbolic Plane}},\ }\href
  {https://doi.org/10.1051/jphyslet:01982004308024900} {\bibfield  {journal}
  {\bibinfo  {journal} {J. Phys. Lett.}\ }\textbf {\bibinfo {volume} {43}},\
  \bibinfo {pages} {249} (\bibinfo {year} {1982})}\BibitemShut {NoStop}%
\bibitem [{\citenamefont {Hobson}\ and\ \citenamefont
  {Nierenberg}(1953)}]{Hobson53}%
  \BibitemOpen
  \bibfield  {author} {\bibinfo {author} {\bibfnamefont {J.~P.}\ \bibnamefont
  {Hobson}}\ and\ \bibinfo {author} {\bibfnamefont {W.~A.}\ \bibnamefont
  {Nierenberg}},\ }\bibfield  {title} {\bibinfo {title} {{The Statistics of a
  Two-Dimensional, Hexagonal Net}},\ }\href
  {https://doi.org/10.1103/PhysRev.89.662} {\bibfield  {journal} {\bibinfo
  {journal} {Phys. Rev.}\ }\textbf {\bibinfo {volume} {89}},\ \bibinfo {pages}
  {662} (\bibinfo {year} {1953})}\BibitemShut {NoStop}%
\bibitem [{\citenamefont {Thorpe}(1981)}]{Thorpe81}%
  \BibitemOpen
  \bibfield  {author} {\bibinfo {author} {\bibfnamefont {M.~F.}\ \bibnamefont
  {Thorpe}},\ }\href@noop {} {\emph {\bibinfo {title} {{Excitations in
  Disordered Systems}}}}\ (\bibinfo  {publisher} {Plenum},\ \bibinfo {address}
  {New-York},\ \bibinfo {year} {1981})\BibitemShut {NoStop}%
\bibitem [{\citenamefont {Koll\'ar}\ \emph {et~al.}(2019)\citenamefont
  {Koll\'ar}, \citenamefont {Fitzpatrick},\ and\ \citenamefont
  {Houck}}]{Kollar19}%
  \BibitemOpen
  \bibfield  {author} {\bibinfo {author} {\bibfnamefont {A.~J.}\ \bibnamefont
  {Koll\'ar}}, \bibinfo {author} {\bibfnamefont {M.}~\bibnamefont
  {Fitzpatrick}},\ and\ \bibinfo {author} {\bibfnamefont {A.~A.}\ \bibnamefont
  {Houck}},\ }\bibfield  {title} {\bibinfo {title} {{Hyperbolic lattices in
  circuits quantum electrodynamics}},\ }\href
  {https://doi.org/10.1038/s41586-019-1348-3} {\bibfield  {journal} {\bibinfo
  {journal} {Nature (London)}\ }\textbf {\bibinfo {volume} {571}},\ \bibinfo
  {pages} {45} (\bibinfo {year} {2019})}\BibitemShut {NoStop}%
\bibitem [{\citenamefont {Koll\'ar}\ \emph {et~al.}(2020)\citenamefont
  {Koll\'ar}, \citenamefont {Fitzpatrick}, \citenamefont {Sarnak},\ and\
  \citenamefont {Houck}}]{Kollar20}%
  \BibitemOpen
  \bibfield  {author} {\bibinfo {author} {\bibfnamefont {A.~J.}\ \bibnamefont
  {Koll\'ar}}, \bibinfo {author} {\bibfnamefont {M.}~\bibnamefont
  {Fitzpatrick}}, \bibinfo {author} {\bibfnamefont {P.}~\bibnamefont
  {Sarnak}},\ and\ \bibinfo {author} {\bibfnamefont {A.~A.}\ \bibnamefont
  {Houck}},\ }\bibfield  {title} {\bibinfo {title} {{Line-graph Lattices:
  Euclidean and Non-Euclidean Flat Bands, and Implementations in Circuit
  Quantum Electrodynamics}},\ }\href
  {https://doi.org/10.1007/s00220-019-03645-8} {\bibfield  {journal} {\bibinfo
  {journal} {Commun. Math. Phys.}\ }\textbf {\bibinfo {volume} {376}},\
  \bibinfo {pages} {1909} (\bibinfo {year} {2020})}\BibitemShut {NoStop}%
\bibitem [{\citenamefont {Bzdu\v{s}ek}\ and\ \citenamefont
  {Maciejko}(2022)}]{Bzdusek22}%
  \BibitemOpen
  \bibfield  {author} {\bibinfo {author} {\bibfnamefont {T.}~\bibnamefont
  {Bzdu\v{s}ek}}\ and\ \bibinfo {author} {\bibfnamefont {J.}~\bibnamefont
  {Maciejko}},\ }\bibfield  {title} {\bibinfo {title} {{Flat bands and
  band-touching from real-space topology in hyperbolic lattices}},\ }\href
  {https://doi.org/10.1103/PhysRevB.106.155146} {\bibfield  {journal} {\bibinfo
   {journal} {Phys. Rev. B}\ }\textbf {\bibinfo {volume} {106}},\ \bibinfo
  {pages} {155146} (\bibinfo {year} {2022})}\BibitemShut {NoStop}%
\bibitem [{\citenamefont {Mosseri}\ \emph {et~al.}(2022)\citenamefont
  {Mosseri}, \citenamefont {Vogeler},\ and\ \citenamefont {Vidal}}]{Mosseri22}%
  \BibitemOpen
  \bibfield  {author} {\bibinfo {author} {\bibfnamefont {R.}~\bibnamefont
  {Mosseri}}, \bibinfo {author} {\bibfnamefont {R.}~\bibnamefont {Vogeler}},\
  and\ \bibinfo {author} {\bibfnamefont {J.}~\bibnamefont {Vidal}},\ }\bibfield
   {title} {\bibinfo {title} {{Aharonov-Bohm cages, flat bands, and gap
  labeling in hyperbolic tilings}},\ }\href
  {https://doi.org/10.1103/PhysRevB.106.155120} {\bibfield  {journal} {\bibinfo
   {journal} {Phys. Rev. B}\ }\textbf {\bibinfo {volume} {106}},\ \bibinfo
  {pages} {155120} (\bibinfo {year} {2022})}\BibitemShut {NoStop}%
\bibitem [{\citenamefont {Boettcher}\ \emph {et~al.}(2020)\citenamefont
  {Boettcher}, \citenamefont {Bienias}, \citenamefont {Belyansky},
  \citenamefont {Koll\'ar},\ and\ \citenamefont {Gorshkov}}]{Boettcher20}%
  \BibitemOpen
  \bibfield  {author} {\bibinfo {author} {\bibfnamefont {I.}~\bibnamefont
  {Boettcher}}, \bibinfo {author} {\bibfnamefont {P.}~\bibnamefont {Bienias}},
  \bibinfo {author} {\bibfnamefont {R.}~\bibnamefont {Belyansky}}, \bibinfo
  {author} {\bibfnamefont {A.~J.}\ \bibnamefont {Koll\'ar}},\ and\ \bibinfo
  {author} {\bibfnamefont {A.~V.}\ \bibnamefont {Gorshkov}},\ }\bibfield
  {title} {\bibinfo {title} {{Quantum simulation of hyperbolic space with
  circuit quantum electrodynamics: From graphs to geometry}},\ }\href
  {https://doi.org/10.1103/PhysRevA.102.032208} {\bibfield  {journal} {\bibinfo
   {journal} {Phys. Rev. A}\ }\textbf {\bibinfo {volume} {102}},\ \bibinfo
  {pages} {032208} (\bibinfo {year} {2020})}\BibitemShut {NoStop}%
\bibitem [{\citenamefont {Urwyler}\ \emph {et~al.}(2022)\citenamefont
  {Urwyler}, \citenamefont {Lenggenhager}, \citenamefont {Boettcher},
  \citenamefont {Thomale}, \citenamefont {Neupert},\ and\ \citenamefont
  {Bzdu\v{s}ek}}]{Urwyler22}%
  \BibitemOpen
  \bibfield  {author} {\bibinfo {author} {\bibfnamefont {D.~M.}\ \bibnamefont
  {Urwyler}}, \bibinfo {author} {\bibfnamefont {P.~M.}\ \bibnamefont
  {Lenggenhager}}, \bibinfo {author} {\bibfnamefont {I.}~\bibnamefont
  {Boettcher}}, \bibinfo {author} {\bibfnamefont {R.}~\bibnamefont {Thomale}},
  \bibinfo {author} {\bibfnamefont {T.}~\bibnamefont {Neupert}},\ and\ \bibinfo
  {author} {\bibfnamefont {T.}~\bibnamefont {Bzdu\v{s}ek}},\ }\bibfield
  {title} {\bibinfo {title} {{Hyperbolic Topological Band Insulators}},\ }\href
  {https://doi.org/10.1103/PhysRevLett.129.246402} {\bibfield  {journal}
  {\bibinfo  {journal} {Phys. Rev. Lett.}\ }\textbf {\bibinfo {volume} {129}},\
  \bibinfo {pages} {246402} (\bibinfo {year} {2022})}\BibitemShut {NoStop}%
\bibitem [{\citenamefont {Gluscevich}\ \emph {et~al.}()\citenamefont
  {Gluscevich}, \citenamefont {Samanta}, \citenamefont {Manna},\ and\
  \citenamefont {Roy}}]{Gluscevich23}%
  \BibitemOpen
  \bibfield  {author} {\bibinfo {author} {\bibfnamefont {N.}~\bibnamefont
  {Gluscevich}}, \bibinfo {author} {\bibfnamefont {A.}~\bibnamefont {Samanta}},
  \bibinfo {author} {\bibfnamefont {S.}~\bibnamefont {Manna}},\ and\ \bibinfo
  {author} {\bibfnamefont {B.}~\bibnamefont {Roy}},\ }\href@noop {} {\bibinfo
  {title} {Dynamic mass generation on two-dimensional electronic hyperbolic
  lattices}},\ \bibinfo {note}
  {\href{https://arxiv.org/abs/2302.04864}{arXiv:2302.04864}}\BibitemShut
  {NoStop}%
\bibitem [{\citenamefont {Nagnibeda}(1999)}]{Nagnibeda99}%
  \BibitemOpen
  \bibfield  {author} {\bibinfo {author} {\bibfnamefont {T.}~\bibnamefont
  {Nagnibeda}},\ }\bibfield  {title} {\bibinfo {title} {{An estimate of
  spectral spectral radii of random walks on surface groups}},\ }\href
  {https://doi.org/10.1007/BF02175833} {\bibfield  {journal} {\bibinfo
  {journal} {J. Math. Sci.}\ }\textbf {\bibinfo {volume} {96}},\ \bibinfo
  {pages} {3542} (\bibinfo {year} {1999})}\BibitemShut {NoStop}%
\bibitem [{\citenamefont {Gou\"ezel}(2015)}]{Gouezel15}%
  \BibitemOpen
  \bibfield  {author} {\bibinfo {author} {\bibfnamefont {S.}~\bibnamefont
  {Gou\"ezel}},\ }\bibfield  {title} {\bibinfo {title} {{A numerical lower
  bound for the spectral radius of random walks on surface groups}},\ }\href
  {https://doi.org/10.1017/S0963548314000819} {\bibfield  {journal} {\bibinfo
  {journal} {Combinator. Probab. Comp.}\ }\textbf {\bibinfo {volume} {24}},\
  \bibinfo {pages} {238} (\bibinfo {year} {2015})}\BibitemShut {NoStop}%
\bibitem [{\citenamefont {Lux}\ and\ \citenamefont {Prodan}()}]{Lux22}%
  \BibitemOpen
  \bibfield  {author} {\bibinfo {author} {\bibfnamefont {F.~R.}\ \bibnamefont
  {Lux}}\ and\ \bibinfo {author} {\bibfnamefont {E.}~\bibnamefont {Prodan}},\
  }\href@noop {} {\bibinfo {title} {{Spectral and combinatorial aspects of
  Cayley-crystals}}},\ \bibinfo {note}
  {\href{https://arxiv.org/abs/2212.10329v4}{arXiv:2212.10329}}\BibitemShut
  {NoStop}%
\bibitem [{\citenamefont {Higuchi}\ and\ \citenamefont
  {Shirai}(2003)}]{Higuchi03}%
  \BibitemOpen
  \bibfield  {author} {\bibinfo {author} {\bibfnamefont {Y.}~\bibnamefont
  {Higuchi}}\ and\ \bibinfo {author} {\bibfnamefont {T.}~\bibnamefont
  {Shirai}},\ }\bibfield  {title} {\bibinfo {title} {{Isoperimetric Constants
  of $(d, f)$-Regular Planar Graphs}},\ }\href
  {https://doi.org/10.4036/iis.2003.221} {\bibfield  {journal} {\bibinfo
  {journal} {Interdiscip. Inf. Sci.}\ }\textbf {\bibinfo {volume} {9}},\
  \bibinfo {pages} {221} (\bibinfo {year} {2003})}\BibitemShut {NoStop}%
\bibitem [{\citenamefont {Cheeger}(1970)}]{Cheeger70}%
  \BibitemOpen
  \bibfield  {author} {\bibinfo {author} {\bibfnamefont {J.}~\bibnamefont
  {Cheeger}},\ }\href@noop {} {\emph {\bibinfo {title} {{Problems in Analysis
  (A Symposium in Honor of S. Bochner)}}}}\ (\bibinfo  {publisher} {Princeton
  University Press},\ \bibinfo {address} {Princeton, NJ},\ \bibinfo {year}
  {1970})\BibitemShut {NoStop}%
\bibitem [{\citenamefont {Urwyler}()}]{Urwyler_thesis}%
  \BibitemOpen
  \bibfield  {author} {\bibinfo {author} {\bibfnamefont {D.~M.}\ \bibnamefont
  {Urwyler}},\ }\href {https://doi.org/10.13140/RG.2.2.34715.34081} {\bibinfo
  {title} {{Hyperbolic Topological Insulator, Master Thesis, University of
  Z\"urich, (2021)}}}\BibitemShut {NoStop}%
\bibitem [{\citenamefont {Chen}\ \emph {et~al.}(2023)\citenamefont {Chen},
  \citenamefont {Brand}, \citenamefont {Helbig}, \citenamefont {Hofmann},
  \citenamefont {Imhof}, \citenamefont {Fritzsche}, \citenamefont
  {Kie{\ss}ling}, \citenamefont {Stegmaier}, \citenamefont {Upreti},
  \citenamefont {Neupert}, \citenamefont {Bzdu\v{s}ek}, \citenamefont
  {Greiter}, \citenamefont {Thomale},\ and\ \citenamefont
  {Boettcher}}]{Chen23}%
  \BibitemOpen
  \bibfield  {author} {\bibinfo {author} {\bibfnamefont {A.}~\bibnamefont
  {Chen}}, \bibinfo {author} {\bibfnamefont {H.}~\bibnamefont {Brand}},
  \bibinfo {author} {\bibfnamefont {T.}~\bibnamefont {Helbig}}, \bibinfo
  {author} {\bibfnamefont {T.}~\bibnamefont {Hofmann}}, \bibinfo {author}
  {\bibfnamefont {S.}~\bibnamefont {Imhof}}, \bibinfo {author} {\bibfnamefont
  {A.}~\bibnamefont {Fritzsche}}, \bibinfo {author} {\bibfnamefont
  {T.}~\bibnamefont {Kie{\ss}ling}}, \bibinfo {author} {\bibfnamefont
  {A.}~\bibnamefont {Stegmaier}}, \bibinfo {author} {\bibfnamefont {L.~K.}\
  \bibnamefont {Upreti}}, \bibinfo {author} {\bibfnamefont {T.}~\bibnamefont
  {Neupert}}, \bibinfo {author} {\bibfnamefont {T.}~\bibnamefont
  {Bzdu\v{s}ek}}, \bibinfo {author} {\bibfnamefont {M.}~\bibnamefont
  {Greiter}}, \bibinfo {author} {\bibfnamefont {R.}~\bibnamefont {Thomale}},\
  and\ \bibinfo {author} {\bibfnamefont {I.}~\bibnamefont {Boettcher}},\
  }\bibfield  {title} {\bibinfo {title} {{Hyperbolic matter in electrical
  circuits with tunable complex phases}},\ }\href
  {https://doi.org/10.1038/s41467-023-36359-6} {\bibfield  {journal} {\bibinfo
  {journal} {Nat. Commun.}\ }\textbf {\bibinfo {volume} {14}},\ \bibinfo
  {pages} {622} (\bibinfo {year} {2023})}\BibitemShut {NoStop}%
\end{thebibliography}%

%

%
\appendix

\section {Size of the clusters as a function of the radius $R$}
\label{app:size}

In this appendix, we provide some recursive formulas that allows one to compute the number of sites of hyperbolic $\{p,3\}$ tilings of radius $R$ used in this paper. Starting from the central site, it is helpful to introduce the notion of shell defined as the set of sites located at a given (graph) distance. By definition, the $R^{th}$ shell corresponds to a radius $R$. 
Let us denote by $n_j$ the total number of sites of the $j^{th}$ shell. and by $d_j$ the number of sites on the $j^{th}$ shell having two neighbors in the $(j+1)^{th}$ shell [the remaining $(n_j-d_j)$ sites  have only one neighbor on the $(j+1)^{th}$ shell]. 

A close inspection of the shell-by-shell growth leads to the following recursive relations
\beqn
d_j&=&2 d_{j-1} - 2 d_{j-r} + d_{j-r-1}, \\
n_j&=&d_j+d_{j-r},
\eeqn
for even $p=2r$, and 
\beqn
d_j&=& 2 d_{j-1} - 2 d_{j-r} + 2 d_{j-r-1}- 2 d_{j-2r}+d_{j-2r-1}, \qquad  \\
n_j&=&d_j + 2 d_{j-r}+  d_{j-2r},
\eeqn
for odd $p=2r+1$. These relations hold for $j \geqslant r$ with the following initial conditions: $d_{j<0}=0$, $d_0=3$, and \mbox{$d_{1\leqslant j \leqslant r-1}=3 \times 2^{j-1}$}.

The total number of sites in a cluster of radius $R$ is finally given by 
\be 
V(R)=1+\sum_{j=1}^R n_j.
\ee
Using these relations, it is straightforward to extract that the asymptotic growth rate  of any hyperbolic $\{p,3\}$ tilings $\lambda_p=\lim_{R \to \infty} \frac{V(R+1)}{V(R)}$. It is given here by the largest nonnegative (Pisot-Vijayaraghavan) root of the polynomial equation
\be 
x^{r+1}-2x^{r}+2x-1=0,
\ee
for even $p=2r$, and 
\be
x^{2r+1}- 2 x^{2r} +2 x^{r+1} - 2 x^r+ 2x -1= 0,
\ee
for odd $p=2r+1$.

This gives the exponential growth  $V(R) \sim \lambda_p^R$ expected for regular hyperbolic tilings. For the limiting case $p=6$  (honeycomb lattice), one gets $\lambda_6=1$, which is reminiscent of a drastically different scaling in the Euclidean plane where $V(R)\sim R^2$. In the large-$p$ limit, one recovers the growth rate, $\lambda_\infty=2$, of the 3-regular Bethe lattice. Remarkably, for $p=8, 10$, one finds the following simple analytical expressions
\beqn
\lambda_8&=& \frac{1}{4} \left(1+\sqrt{13}+\sqrt{2 \sqrt{13}-2}\right)\simeq 1.72208,\\
\lambda_{10}&=&\frac{1}{2}\left(1+\sqrt{2} +\sqrt{2 \sqrt{2}-1}\right) \simeq 1.88320.
\eeqn
These values are in agreement with the numerical results given in the Supplemental Material of Ref.~\cite{Chen23}.  As can be easily checked, $\lambda_p$ is a monotonically increasing function of \mbox{$p\geqslant 6$}. 

\section {Continued-fraction coefficients}
\label{app:coeff}
In this Appendix, we give the coefficients $(a_n,b_n)$ for the $\{p,3\}$ tilings considered in this paper as well as for the $\{8,8\}$ tiling. These coefficients are all rational numbers but, for the sake of clarity,  we only give the first ten coefficients in this form.

\begin{table*}[h]
\center
\begin{tabular}{| c | c | c | c | }
\hline
 & $\{ 7,3 \}$ & \{ 9,3 \} & \{ 11,3 \}  \\
\hline
\hline  
$a_1$ & 0 & 0 & 0 \\
\hline  
$a_2$ & 0 & 0 & 0  \\
\hline  
$a_3$ & 0 & 0 & 0  \\
\hline  
$a_4$ & -$\frac{1}{2}$ & 0 & 0  \\
\hline  
$a_5$ & -$\frac{1}{14}$ & -$\frac{1}{4}$ & 0  \\
\hline  
$a_6$ & -$\frac{109}{287}$ & -$\frac{1}{124}$ & -$\frac{1}{8}$  \\
\hline  
$a_7$ & -$\frac{895}{20\,869}$ & -$\frac{2\,426}{28\,799}$ &-$\frac{1}{1\,016}$  \\
\hline  
$a_8$ & -$\frac{215\,701}{14\,48\,614}$ &-$\frac{2\,977\,042}{24\,348\,161}$  &-$\frac{68\,596}{2\,032\,127}$  \\
\hline  
$a_9$ & -$\frac{22\,367\,419}{92\,514\,922}$ &  -$\frac{2\,720\,237\,055}{40\,071\,097\,354}$ &-$\frac{1\,190\,359\,196}{31\,846\,454\,279}$  \\
\hline  
$a_{10}$ & -$\frac{13\,311\,270\,229}{86\,595\,884\,905}$ &-$ \frac{20\,055\,768\,316\,639}{259\,125\,041\,818\,854}$ & -$\frac{21\,445\,227\,755\,471}{483\,141\,503\,018\,839}$  \\
\hline  
$a_{11}$ & -0.2002967411& -0.0645775422  & -0.0357443812  \\
\hline  
$a_{12}$ & -0.1696004570 & -0.1058977166 & -0.0383582342  \\
\hline  
$a_{13}$ & -0.1816966879 & -0.0657104328 & -0.0348093723  \\
\hline  
$a_{14}$ & -0.1692155243 & -0.0795238494 & -0.0319445279  \\
\hline  
$a_{15}$ & -0.1890266285 & -0.0878232196 & -0.0442313413  \\
\hline  
$a_{16}$ & -0.1824985928 & -0.0785198490 & -0.0346354601  \\
\hline  
$a_{17}$ & -0.1710620428 & -0.0806426945 & -0.0361022603  \\
\hline  
$a_{18}$ & -0.1852717728& -0.0801376552 & -0.0362444720  \\
\hline  
$a_{19}$ & -0.1752233562 & -0.0817817362 & -0.0377254575  \\
\hline  
$a_{20}$ & -0.1832919042 & -0.0808816316 & -0.0367628802  \\
\hline  
$a_{21}$ & -0.1755484537 & -0.0804796200 & -0.0370226959  \\
\hline  
$a_{22}$ & -0.1823834605 & -0.0805314355 & -0.0366759599  \\
\hline  
$a_{23}$ & -0.1789067207 & -0.0813731403 & -0.0364735833  \\
\hline  
$a_{24}$ & -0.1778987384 & -0.0809167668 & -0.0371592471  \\
\hline  
$a_{25}$ & -0.1803519218 & -0.0805287669 & -0.0369329370  \\
\hline  
$a_{26}$ & -0.1807663195 & -0.0808180145 & -0.0367983443  \\
\hline  
$a_{27}$ & -0.1773146456 & -0.0810638822 & -0.0367319221  \\
\hline  
$a_{28}$ & -0.1804581485& -0.0808744295 & -0.0368556230  \\
\hline  
$a_{29}$ & -0.1802893026 & -0.0806615422 & -0.0368804917  \\
\hline  
$a_{30}$ & -0.1782479737 & -0.0808573363 & -0.0368682992  \\
\hline  
$a_{31}$ & -0.1797527809 & -0.0810001089 &   \\
\hline  
$a_{32}$ & -0.1804963609 & -0.0807915986 &  \\
\hline  
$a_{33}$ & -0.1783586603 &  &   \\
\hline  
$a_{34}$ & -0.1799096739 &  &   \\
\hline  
$a_{35}$ & -0.1798647291 &  &   \\
\hline  
$a_{36}$ & -0.1790760687 &  &   \\
\hline  
$a_{37}$ & -0.1795811934 &  &   \\
\hline  
$a_{38}$ & -0.1797409614 &  &   \\
\hline  
$a_{39}$ & -0.1792827917 &  &   \\
\hline  
$a_{40}$ & -0.1795777244 &  &   \\
\hline  
$a_{41}$ & -0.1795289697 &  &   \\
\hline  
$a_{42}$ & -0.1794620439 &  &   \\
\hline
\end{tabular}
\caption{List of coefficients $a_n$ for all $\{p,3\}$ tilings studied in this paper. For bipartite tilings (even $p$),and $a_{n\geqslant 1}=0$.}
\label{tab:conder}
\end{table*}

%
\begin{table*}[h]
\center
\begin{tabular}{| c | c | c | c | c | c | c |}
\hline
& \{ 7,3 \} & \{ 8,3 \} & \{ 9,3 \} & \{ 10,3 \} & \{ 11,3 \} & \{ 12,3 \} \\
\hline
\hline  
$b_{1}$ & 3 & 3 & 3 & 3 & 3 & 3  \\
\hline
$b_{2}$ &  2 & 2 & 2 & 2 & 2 & 2  \\
\hline
$b_{3}$ &  2 & 2 & 2 & 2 & 2 & 2  \\
\hline
$b_{4}$ &  $\frac{7}{4}$ & $\frac{5}{2}$ & 2 & 2 & 2 & 2  \\
 \hline
$b_{5}$ & $\frac{82}{49}$ & $\frac{19}{10}$  & $\frac{31}{16}$ & $\frac{9}{4}$  & 2 & 2  \\
 \hline
$b_{6}$ & $\frac{3\,563}{1\,681}$ & $\frac{192}{95}$ & $\frac{1\,858}{961}$ & $\frac{71}{36}$  & $\frac{127}{64}$ & $\frac{17}{8}$  \\
\hline
$b_{7}$ & $\frac{466\,744}{259\,081}$ & $\frac{1\,435}{608}$ & $\frac{1\,624\,958}{863\,041}$ & $\frac{1\,298}{639}$  & $\frac{32\,002}{16\,129}$ & $\frac{271}{136}$  \\
\hline
$b_{8}$ & $\frac{16\,546\,063}{8\,099\,716}$ & $\frac{53\,675}{27\,552}$ & $\frac{1\,420\,353\,674}{686\,911\,681}$ & $\frac{93\,114}{46\,079}$  &$\frac{505\,530\,866}{256\,032\,001}$ & $\frac{9\,318}{4\,607}$  \\
\hline
$b_{9}$ & $\frac{3\,790\,751\,045}{2\,113\,410\,098} $& $\frac{5\,117\,344}{2\,432\,325}$ &$ \frac{4\,442\,005\,081\,431}{2\,337\,553\,556\,836}$ & $\frac{1\,023\,678}{479\,611}$  & $\frac{7\,768\,506\,013\,282}{3\,961\,210\,497\,841}$ & $\frac{2\,549\,218}{1\,262\,589}$  \\
\hline
$b_{10}$ & $\frac{27\,441\,726\,460\,437}{14\,192\,886\,254\,450}$ & $\frac{996\,022\,307}{451\,765\,525}$ & $\frac{56\,260\,939\,420\,701\,038}{28\,724\,812\,358\,313\,681}$ & $\frac{21\,518\,893}{10\,654\,902}$  &$ \frac{11\,925\,020\,979\,148\,0311}{58\,927\,873\,705\,910\,881}$ & $\frac{703\,998\,380}{349\,317\,843}$  \\
\hline
$b_{11}$ & 1.9099614142 & 2.0409742217 & 1.9901940863 & 2.0437041571  & 1.9696415108 & 2.0511292014  \\
\hline
$b_{12}$ & 1.8980216501 & 2.1147900908 & 1.9488793608 & 2.0424319634  & 1.9825865391 & 2.0192825585  \\
\hline
$b_{13}$ & 1.9188757385 & 2.1343590649 & 1.9603634536 & 2.0804121861  & 1.9837731591 & 2.0266312300  \\
\hline
$b_{14}$ & 1.9076043684 & 2.0929965018 & 1.9556323717 & 2.0428949588  & 1.9890943205 & 2.0239897720  \\
\hline
$b_{15}$ & 1.8984122342 & 2.1117327570 & 1.9689855709 & 2.0489672866  & 1.9842666406 & 2.0233232583  \\
\hline
$b_{16}$ & 1.9014941806 & 2.1092519737 & 1.9562792157 & 2.0493838687  & 1.9859259226 & 2.0321707408  \\
\hline
$b_{17}$ & 1.9227626558 & 2.1128371617 & 1.9593538591 & 2.0606714712  & 1.9842298525 & 2.0255750100  \\
\hline
$b_{18}$ & 1.8964658469 & 2.1080093699 & 1.9622805785 & 2.0511933622  & 1.9831821766 & 2.0267955811  \\
\hline
$b_{19}$ & 1.9057267568 & 2.1063772837 & 1.9608916272 & 2.0513327471  & 1.9874982510 & 2.0257927666  \\
\hline
$b_{20}$ & 1.9153463807 & 2.1128820963 & 1.9604032064 & 2.0514394969  & 1.9848285336 & 2.0256152673  \\
\hline
$b_{21}$ & 1.8989460074 & 2.1101619721 & 1.9600571477 & 2.0545544548  & 1.9847441368 & 2.0277187917  \\
\hline
$b_{22}$ & 1.9083032359 & 2.1063672332 & 1.9606957617 & 2.0531377036  & 1.9846896144 & 2.0266121387  \\
\hline
$b_{23}$ & 1.9099285244 & 2.1106767717 & 1.9611107232 & 2.0524104665  & 1.9853636852 & 2.0267155107  \\
\hline
$b_{24}$ & 1.9043990627 & 2.1112849502 & 1.9604826517 & 2.0521731483  & 1.9851526874 & 2.0263860553  \\
\hline
$b_{25}$ & 1.9060675304 & 2.1075971495 & 1.9600890915 & 2.0529601604  & 1.9851660959 & 2.0262903312 \\
\hline
$b_{26}$ & 1.9084876568 & 2.1091525930 & 1.9609144549 & 2.0532068120  & 1.9850303328 & 2.0267658961  \\
\hline
$b_{27}$ & 1.9059486122 & 2.1111627628 & 1.9608480799 & 2.0528183119  & 1.9849405419 & 2.0266632108  \\
\hline
$b_{28}$ & 1.9060851409 & 2.1086943848 & 1.9603493231& 2.0525196876  & 1.9851286602 & 2.0266541652  \\
\hline
$b_{29}$ & 1.9075925091 & 2.1088975541 & 1.9605057348 & 2.0526635295  & 1.9851666383 & 2.0265531284  \\
\hline
$b_{30}$ & 1.9061094109 & 2.1102067643 & 1.9607455795 & 2.0529841666  & 1.9850980048 & 2.0264998061  \\
\hline
$b_{31}$ & 1.9070439609 & 2.1096212693 & 1.9606731972 & 2.0529054816  &  &   \\
\hline
$b_{32}$ & 1.9060671213 & 2.1089557441 & 1.9605012778 &   &  &   \\
\hline
$b_{33}$ & 1.9071946002 & 2.1095759718 &  &   &  &   \\
\hline
$b_{34}$ & 1.9066326631 & 2.1099008138 &  &   &  &   \\
\hline
$b_{35}$ & 1.9062481065 & 2.1092213222 &  &   &  &   \\
\hline
$b_{36}$ & 1.9068805138 & &  &   &  &   \\
\hline
$b_{37}$ & 1.9070449307 &  &  &   &  &   \\
\hline
$b_{38}$ & 1.9059497758 &  &  &   &  &   \\
\hline
$b_{39}$ & 1.9071011410 &  &  &   &  &   \\
\hline
$b_{40}$ & 1.9067937263 &  &  &   &  &   \\
\hline
$b_{41}$ & 1.9063003478 & &  &   &  &   \\
\hline
$b_{42}$ & 1.9068134885 & &  &   &  &   \\
\hline
\end{tabular}
\caption{List of coefficients $b_n$ for all $\{p,3\}$ tilings studied in this paper. In the large-$p$ limit these tilings converge towards the 3-regular Bethe lattice for which one has $b_1=3$ and $b_{n\geqslant 2}=2$.}
\label{tab:coeffs_b_p_3}
\end{table*}

%
%
\begin{table*}[t]
\center
\begin{tabular}{| c | c | c | }
\hline
$n$  &$b_n$ & $\langle H^{2n} \rangle$ \\
\hline
1 &8&8\\
\hline
2  &7& 120\\
\hline
3 &7&2\,192\\
\hline
4 &$\frac{345}{49}$ &44\,264\\
\hline
5  & $\frac{39\,607}{5\,635}$&950\,608\\
\hline
6  &$\frac{32\,015\,739}{4\,554\,805}$& 21\,288\,912\\
\hline
7&$\frac{3\,819\,904\,499\,705}{543\,448\,874\,817}$&491\,515\,088\\
\hline
8  &$\frac{457\,663\,490\,414\,626\,565}{65\,109\,351\,624\,213\,411}$&11\,614\,244\,072\\
\hline
9  &$\frac{385\,418\,200\,444\,183\,773\,404\,967}{54\,831\,603\,309\,520\,014\,006\,895}$&279\,495\,834\,368\\
\hline
10 & $\frac{6\,844\,506\,818\,384\,509\,461\,062\,609\,435\,843}{973\,735\,600\,854\,857\,922\,718\,228\,679\,945}$&6\,826\,071\,585\,040\\
\hline
{\red 11} & \red{$\frac{854\,384\,354\,399\,029\,778\,591\,853\,594\,278\,622\,479\,055}{121\,549\,244\,543\,021\,810\,945\,136\,773\,063\,596\,240\,213}$}& \red{ 168\,755\,930\,104\,880}\\
\hline
{\red 12} & \red{$\frac{249\,887\,384\,305\,886\,872\,771\,075\,994\,660\,075\,645\,942\,846\,356\,081}{35\,550\,298\,162\,871\,632\,768\,930\,141\,283\,569\,631\,274\,191\,260\,129}$}& \red{4\,214\,946\,994\,935\,248}\\
\hline

\end{tabular}
\caption{List of the first exact coefficients $b_n$ for the $\{8,8\}$ tiling computed on a cluster of radius $R=10$ with \mbox{$V=369\, 256\, 049$} sites. As a bipartite tiling, $a_{n \geqslant 1}=0$. In the rightmost column, we also give the exact even moments of the LDOS (odd moments vanish since the tiling is bipartite). These moments are computed by considering the LDOS associated with the first site of the chain whose hopping terms are given by $\sqrt{b_n}$ (see Ref.~\cite{Haydock75}). We also added two more moments (last two rows), provided by S. Gou\"ezel using a completely different approach based on word enumerations~\cite{Gouezel15}, which confirm the quick convergence of the $b_n$'s. The first eight moments can also be found in Ref.~\cite{Lux22}.}
\label{tab:coeffs_b_8_8}
\end{table*}
%
%

\end{document}